\documentclass{aa}
\usepackage{natbib}
\bibpunct{(}{)}{;}{a}{}{,}
\usepackage{graphicx}
\usepackage[percent]{overpic}

\begin{document}
\authorrunning{Chatzistergos et al.}
\titlerunning{GSN: Ultimate backbone method}
\title{New reconstruction of the sunspot group numbers since 1739 using direct calibration
 and `backbone' methods}
\author{Theodosios Chatzistergos\inst{1}, 
Ilya G. Usoskin\inst{2,3},  
Gennady A. Kovaltsov\inst{4}, 
Natalie A. Krivova\inst{1}, 
Sami K. Solanki\inst{1,5}}
\offprints{Theodosios Chatzistergos  \email{chatzistergos@mps.mpg.de}}
\institute{Max-Planck-Institut f\"{u}r Sonnensystemforschung, Justus-von-Liebig-weg 3,	37077 G\"{o}ttingen, Germany
\and Space Climate Research Unit, University of Oulu, FIN-90014, Finland
\and Sodankyl\"a Geophysical Observatory, University of Oulu, FIN-90014, Finland
\and Ioffe Physical-Technical Institute, St.Petersburg, RU-194021 Russia  \and School of Space Research, Kyung Hee University, Yongin, Gyeonggi 446-701, Republic of Korea}
\date{}

\abstract
{The group sunspot number (GSN) series constitute the longest instrumental astronomical database providing
 information on solar activity. 
This database is a compilation of observations by many individual observers, and their inter-calibration has usually been performed using linear rescaling. There are multiple published series that show different long-term trends for solar activity.}
{We aim at producing a GSN series, with a non-linear non-parametric calibration. 
The only underlying assumptions are that the differences between the various series are due to different acuity thresholds
of the observers, and that the threshold of each observer remains constant throughout the observing period.}
{We used a daisy chain process with backbone (BB) observers and calibrated all overlapping observers to them. 
We performed the calibration of each individual observer with a probability distribution function (PDF) matrix
 constructed considering all daily values for the overlapping period with the BB. 
The calibration of the BBs was carried out in a similar manner. 
The final series was constructed by merging different BB series. 
We modelled the propagation of errors straightforwardly with Monte Carlo simulations.
A potential bias due to the selection of BBs was investigated and the effect was shown to lie
 within the 1$\sigma$ interval of the produced series. 
The exact selection of the reference period was shown to have a rather small effect on our calibration as well.}
{The final series extends back to 1739 and includes data from 314 observers. 
This series suggests moderate activity during the 18th and 19th century, which is significantly lower than the high level of solar activity
 predicted by other recent reconstructions applying linear regressions.}
{The new series provides a robust reconstruction, based on modern and non-parametric methods,
 of sunspot group numbers since 1739, and it confirms the existence of the modern grand maximum of solar activity in
 the second half of the 20th century.}
\maketitle

\section{Introduction}

Observations of sunspots on the solar disc have been performed regularly since the advent
 of telescopes in the early 17th century.
These measurements constitute the longest ongoing observational programme in astrophysics,
 providing important insights into solar activity and variability on centennial timescales.

However, these observations have been carried out by different people, with different instruments,
 at various locations. In some cases observations were taken
 for a different purpose but were also later used to define sunspot numbers.
The definition of a sunspot group might have changed with time, gaps exist within the series of individual
 observers, and the various series do not necessarily all overlap with each other.
Even for the same observer, the quality of the record may vary with time owing to, for example gaining experience,
 ageing of the observer (e.g. deteriorating eyesight), change of instrumentation, or varying conditions
 at the observing location.
There have been several attempts to harmonize these measurements and to produce a homogeneous composite series.
The first effort was made by Rudolf Wolf from Z\"urich who introduced the Wolf sunspot number (WSN) in 1848
 \citep[][continued and updated as the international sunspot number, ISN]{wolf_mittheilungen_1850}, given by the formula
\begin{equation}
R_s=k(10\,G+S),
\label{Eq:WSN}
\end{equation}
where $k$ is a weighting factor to normalize the various observers with each other, $S$ the number of sunspots,
 and $G$ the number of sunspot groups. 
It is important that, for the sake of homogeneity, data from only one primary observer were used for each day.
If the data from the primary observer were not available for a given day, data from the secondary, tertiary, etc.,
 observer were used, but only one observation was used per day, ignoring all other available data.
The original records and notebooks of Wolf are not readily available now, implying that WSN cannot be re-constructed from
 scratch.
This series contains annual values back to 1700, while monthly and daily values go back to 1749 and 1818, respectively.
Since 1981 the WSN/ISN series has been synthesized by the Royal Observatory of Belgium \citep{clette_wolf_2007}, adapted to
 include all available observers for each day, rather than only the primary observer.
The WSN/ISN series has been recently updated as version 2.0 by correcting for some proposed inhomogeneities \citep{clette_revisiting_2014}.

More than a century after the work by Wolf, \cite{hoyt_group_1998} introduced the group sunspot number (GSN) series (HoSc98, hereafter),
 which is based on the number of sunspot groups only, neglects individual spots and includes data from all observers on the same day.
The daily GSN is defined as
\begin{equation}
R_g=\frac{12.08}{N} \sum_i k_i G_i,
\end{equation}
where $k_i$ is the individual correction factor of the $i$-th observer, $G_i$ is the GSN reported by the $i$-th observer,
 $N$ is the total number of observers on the given day, and the constant 12.08 was introduced to match the average level of
 $R_g$ to that of $R_s$ over the period 1874--1976.
The GSN series was designed to be more robust than WSN/ISN since it only considers sunspot groups and
 reduces uncertainties in the counts of individual sunspots.
In addition, the GSN series includes a much greater number of raw data than WSN and is extended further back in time to 1610.
An important advantage is that for the GSN series, a complete database of the raw data
 \citep[published as][and revised recently by \citealt{vaquero_revised_2016}]{hoyt_group_1998} is available, which makes it possible
 to reconstruct the entire series from scratch.

The homogenization and cross-calibration of the data recorded by earlier observers was always performed through a
 daisy-chaining sequence of linear scaling normalization of the various observers, using the $k-$factors.
This means that starting with a reference observer, the $k-$factors are derived for overlapping observers.
 The latter data are in turn used as the reference for the next overlapping observers, etc.
As is apparent, this leads to error accumulation in time when moving further away from the reference observer.

It has become obvious that the old series need to be revised because of the new-found data and
 the outdated methodology based on constant $k-$factors. The issue with such methods is twofold. Firstly, such methods assume that counts by two observers are proportional to each other, which is generally not correct. Secondly, the $k-$ factors are assumed to be constant for the entire operational period of each observer, whereas in reality the acuity of the observers and sensitivity of the instruments may vary with time.
A dedicated activity of the research community \citep{clette_revisiting_2014} has led to several new sunspot series
 discussed below.

\citet[][ClLi16, hereafter]{cliver_discontinuity_2016} have attempted to revise the GSN series using essentially the same methodology
 as \citet{hoyt_group_1998}. 
They claim, however, that the earlier part of the Royal Greenwich Observatory (RGO hereafter) data
 (i.e. 41 years before 1915) might suffer from uneven quality owing to the purported learning curve process. 
Therefore, they corrected the GSN values over this period by normalizing them to the data by Wolfer using a second degree polynomial fit.
The inhomogeneity of the early RGO data is still a matter of debate, however.
Other studies did not find any extensive problem with RGO data:
 \citet{sarychev_comparison_2009}, \citet{clette_revisiting_2014}, and \citet{lockwood_assessment_2016}
 reported as potentially problematic periods before 1880, 1900, and 1877, respectively, while data from \citet{aparicio_sunspot_2014} and \citet{carrasco_forty_2013} do not exhibit any apparent trend with respect to RGO data after $\sim$1885 and 1890, respectively.
Thus, the period of 1874--1915 used by ClLi16 to `recalibrate' the RGO dataset is not
 well defined.
The ClLi16 series covers the period 1841--1980 and yields the highest level of sunspot activity in
 the mid-19th century among all available reconstructions.

\citet[][SvSc16, hereafter]{svalgaard_reconstruction_2016} also used the method of daisy-chaining $k-$factors.
 But these authors introduced five key observers (called `backbones', BB hereafter) to calibrate each overlapping secondary observer to these BBs.
 Thus, they seemingly reduced the number of daisy-chain steps
 	because some daisy-chain links are moved
 	into the BB compilation rather than being eliminated.
The problem with this method is that most of the BB observers did not overlap with each other. Thus their inter-calibration was
 performed via series extended using secondary observers with lower quality and poorer statistics.
In the end, this introduces even more daisy-chain steps, since each BB observer is normalized to the neighbouring observer using a three-step
 procedure.
The SvSc16 series also reduced the number of sunspot groups after 1940 by 7\% to take into account the possible
  effect of the introduction of the Waldmeier classification of sunspot groups \citep{waldmeier_uber_1939}.
However \cite{lockwood_tests_2016-4,lockwood_tests_2016-3} have questioned the necessity for such a correction for the GSN.
The SvSc16 series covers the period 1610--2015 and suggests a rather high level of solar activity in the 18th
 and, especially, 17th centuries.

All of these sunspot number series used calibration methods based on the linear scaling regression to derive constant $k-$factors.
However, this linear $k-$factor method has been demonstrated to be unsuitable for such studies \citep{lockwood_tests_2016-1,usoskin_dependence_2016,usoskin_new_2016}, leading to errors in the reconstructions that employ them.

An alternative method was proposed by \citet[][UEA16, hereafter]{usoskin_new_2016}, who calibrated each observer directly to the reference
 dataset, avoiding the daisy chain and error accumulation.
The method is based on comparison of the active day fraction statistics of an observer with that in the reference dataset (RGO data
 for the period 1900--1976).
The quality of each observer is characterized by the acuity observational threshold so that the observer is assumed
 to miss all sunspot groups that are smaller than this threshold, and to report all sunspot groups that are larger than this threshold.
The acuity threshold for each observer is found by matching their active day fraction statistic with that of an artificially
  created reference dataset.
The UEA16 series covers the period 1749--1995 and yields a moderate level of sunspot activity in the 18th and 19th centuries,
 lying between the HoSc98 and SvSc16 series.

Another revision of the GSN series was carried out by \cite{lockwood_centennial_2014} who corrected it for some apparent 
 inhomogeneities.
However, since this study is close to the HoSc98 series, we do not consider it separately here.  

Thus, presently there are a number of sunspot reconstructions using different methods of calibration and yielding
 results that are inconsistent with each other.
The most critical implication of these series is that they yield different long-term trends for the activity of the Sun
 \citep{lockwood_assessment_2016,kopp_impact_2016}.
Over the 19th and 20th centuries, ClLi16 and SvSc16 show no trend, while HoSc98 and UEA16 show an increase in solar activity.

In an attempt to bridge the methodologies underlying previous studies and present more accurate error estimates,
 we present here a recalibration
 of the GSN data using an amendment of the most direct non-parametric calibration method described in \cite{usoskin_dependence_2016}.
Similarly to SvSc16, we incorporate BB observers.
However, the calibration of overlapping observers is performed
 with a non-linear non-parametric probability distribution function (PDF) derived from sunspot group counts
 for days when two observers overlap.
This allows us to account for the error propagation in a straightforward manner.
Calibration of the different resulting BB series is achieved with daisy chaining.

The data we use are introduced in Sec. \ref{s:data}.
The procedure, including information about all individual BB observers and their processing is described in Sec. \ref{s:method}.
Our composite series is presented and compared with other existing series in Sec. \ref{s:results}, where we also discuss
 the stability of our method and potential problems of our series. We summarize our results in Sec. \ref{s:conclusions}.

\section{Data}
\label{s:data}
We employ the database\footnote{Available at \url{http://haso.unex.es/?q=content/data}} of the sunspot group numbers
 recorded by individual observers that was recently published by \cite{vaquero_revised_2016} as an update of the HoSc98 database.
Observers are uniquely identified by their identification number in the database. 
Here we use these identification numbers as well.

We apply the following filters to these data:
\begin{itemize}

\item
Data by \textit{Wolfer} (1880--1928, id 338) were merged with those by \textit{Billwiller and Wolfer} (1876--1879, id 335).
The two series were combined together to a single series, since they do not directly overlap. The two series differ in that the former includes observations solely by Wolfer, while the latter includes observations made by both Wolfer and Billwiller.
By merging these two series together, we can increase the length of the Wolfer series and its overlap with observations by Schmidt.

\item
Data from \textit{Flaugergues, H., Aubenas} (1794--1795, id 22) were also merged with those from \textit{Flaugergues, H., Viviers} (1788--1830, id 227) using the same procedure.
These two datasets were obtained by the same observer, Flaugergues, who performed the bulk of his observations in Viviers, Ard\'eche,
 but who relocated to Aubenas for a period of about two years.

The dataset from Aubenas contains merely 91 observations for these two years, a period of otherwise sparse observations (we have only
 nine records from all other observers used here).
The overlap of the observations of Flaugergues from Aubenas to other observers is less than three days and does not provide adequate statistics
 to properly calibrate this series.
Considering that the two locations are close to each other in the south of France, we make the assumption that the observing
 conditions were not significantly different. This enables us to merge the two Flaugergues series. 
 Furthermore, because of the poor overlap with other series, inclusion of these data does not affect the rest of our series.
\end{itemize}

The HoSc98, ClLi16, SvSc16, and UEA16 series were downloaded from the SILSO\footnote{\url{http://www.sidc.be/silso/groupnumberv3}} (Royal Observatory of Belgium) website.

\section{Calibration process}
\label{s:method}
\subsection{Algorithm and primary observers}
We have developed an automated algorithm to perform the calibration of sunspot records by individual observers which includes the following steps:
\begin{itemize}
\item
First, we selected primary BB observers who provided long and high-quality observations.
\item
Next, we calibrated the data from all other observers, denoted as secondary observers hereafter, to the primary BB observers
 using periods of overlapping observations (sufficient overlap is required, see Section~\ref{s:mat}), and produced the `BB series', which are
 composites of data from the BB observer and all other observers calibrated to him/her.
\item
Individual BB series were cross-calibrated to each other, using the daisy-chain procedure.
\item
Finally, the composite series of daily GSN was constructed by averaging the calibrated BB series.
\end{itemize}
The calibration was carried out using a direct non-parametric method to a single reference dataset with a
 straightforward propagation of errors. No regression was used and the acuity of the observers was assumed constant over their entire observing life. The method is described in detail in Sec. \ref{sec:secondaryobservers}

The selected sequence of the primary BB observers is Kanzelh\"ohe, RGO, Wolfer, Schmidt, Schwabe,
 Flaugergues, and Horrebow (see Table \ref{table:backbones}).
The BB observers were selected to be those with sufficiently long observational records of high quality.
We also used Schubert, Zucconi, and Hagen as stand-alone BBs.
Because of the lacking bridge in the data in the middle of the 18th century, we were unable to directly calibrate these three observers
 to a single observer acting as a BB. Thus we did this by the extended statistics of the calibrated BB series.
These observers are important since they cover periods over the 18th century when no other data are available.
Our reference observer is RGO (but restricted to the period between 1900--1976) and all other
 BB series were calibrated to the level of RGO.

All data from RGO prior to 1900 were ignored when considering the primary BB observer because of the disputed inhomogeneity,
 as discussed in the Introduction.
We discuss the effect of this decision on our calibration in Sect. \ref{sec:rgoquality}.

\begin{figure}
\centering
\includegraphics[width=1\linewidth,bb=113 491 491 751, clip=true]{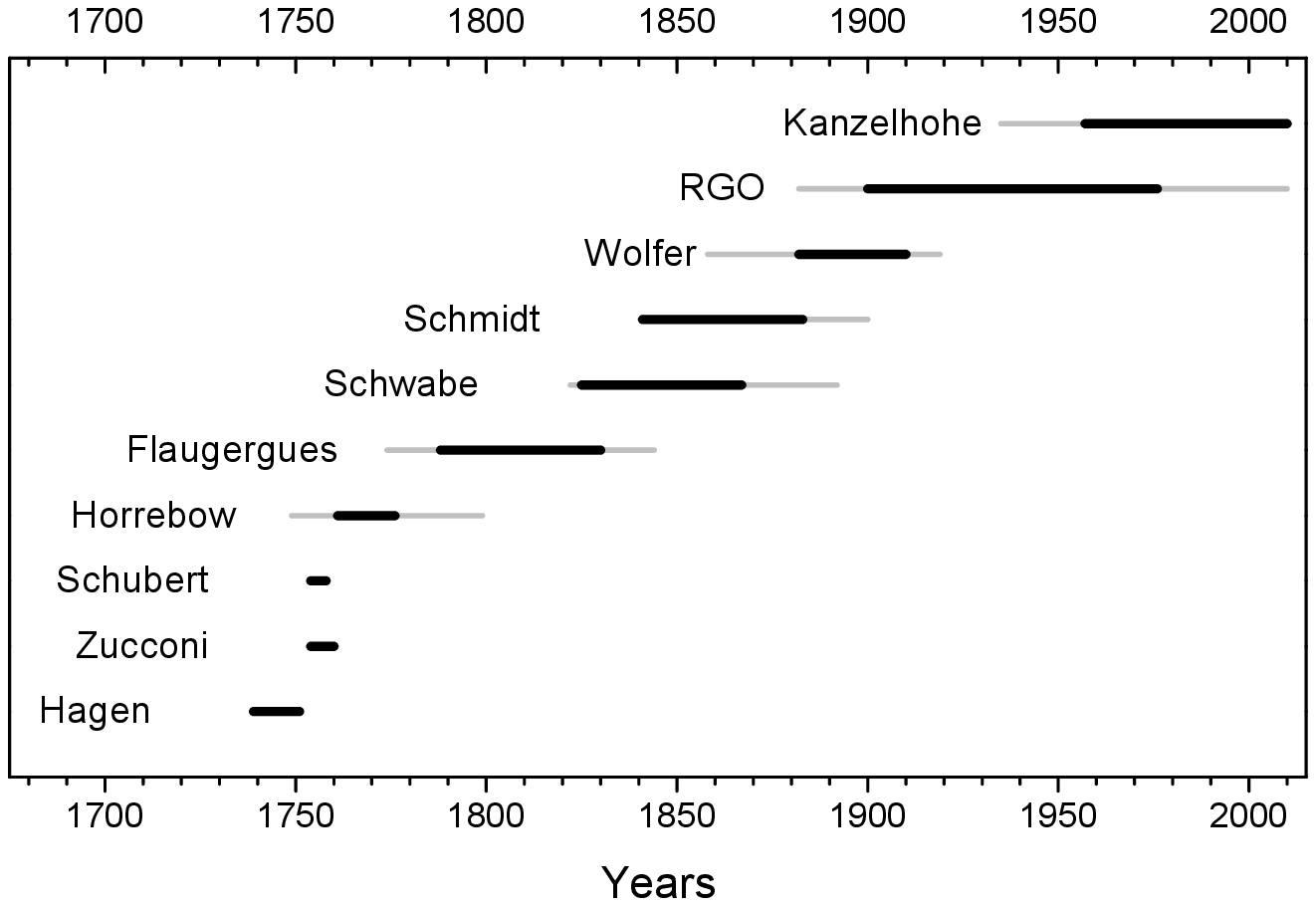}
\caption{Temporal coverage by the BBs used here.
 Solid black lines represent the primary BB observers, while grey lines depict the extension of the
  BBs using calibrated secondary observers.}
\label{fig:BB}
\end{figure}

\subsection{Secondary observers}
\label{sec:secondaryobservers}
Each BB series was also filled with all available secondary observers calibrated to the primary BB observers.
As secondary observers we selected all the observers that have at least one nominal year of overlap with the primary BB observer.
To avoid a distortion of statistics, each observer was included only in one BB.
The assignment of observers to the BBs was made based on the length of the overlapping period and by trying to match observers
 with comparable quality BB observers.
The only two successive BB observers whose observations do not overlap in time are Horrebow and Flaugergues. 
The bridging was made using Staudacher data.
In this case, we chose Horrebow as the BB over Staudacher, because he observed more frequently and the data are of higher quality.
Unfortunately, we were not able to go further back in time than Hagen (1739), because of the very sparse observations over this period with
 no observer making observations both before and after 1739 with adequate data to perform the calibration.
Table \ref{table:backbones} and Figure~\ref{fig:BB} provide key information about the BB observers and series.

\begin{figure}
\centering
\includegraphics[width=1\linewidth,bb=58 144 529 561, clip=true]{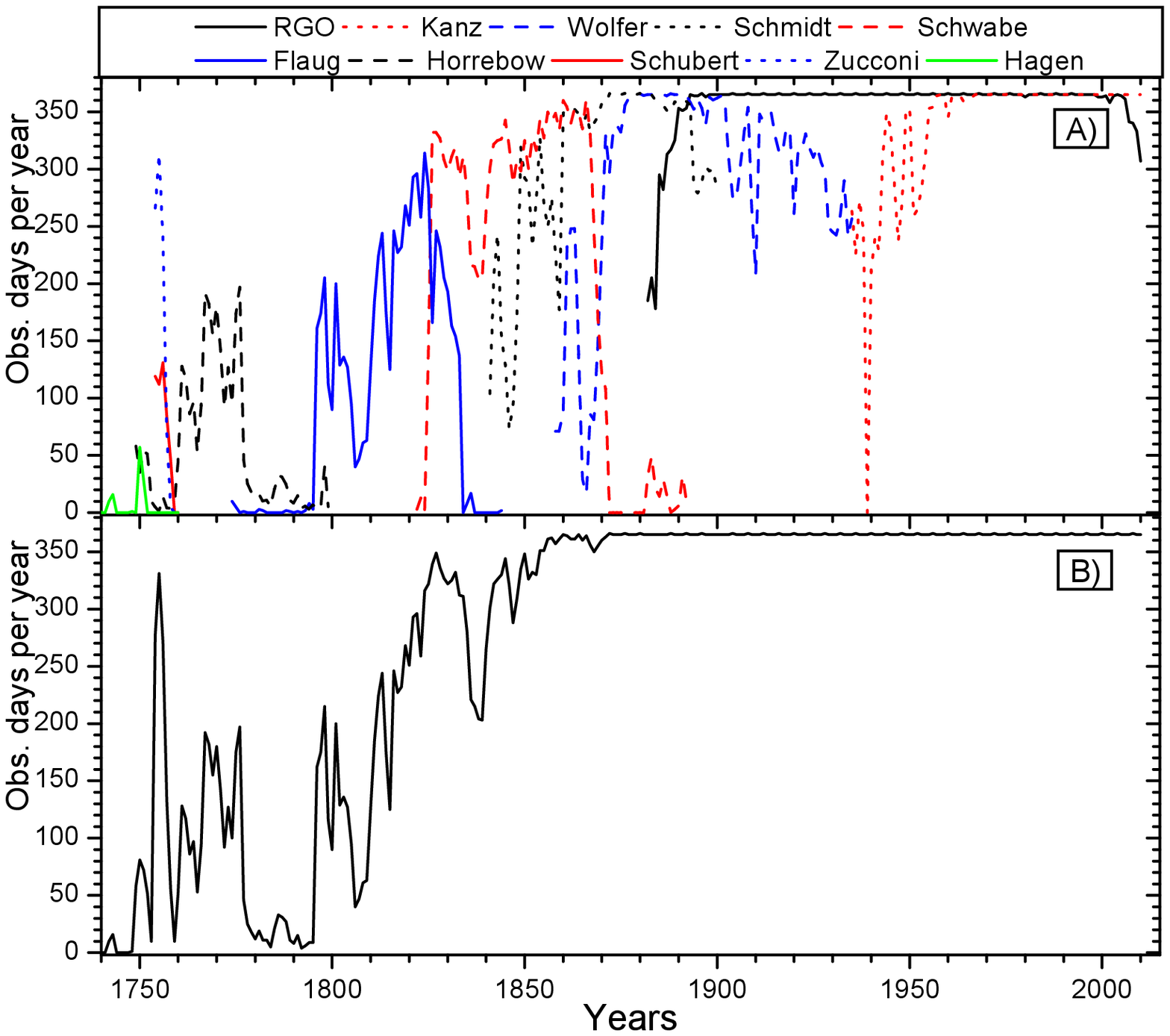}
\caption{Annual coverage (number of observational days per year) by the different BB series (coloured curves in panel a)
 and by our final composite series (panel b).}
\label{fig:numberofdays}
\end{figure}

All the observers we used for various BBs are listed in Tables \ref{table:rgobackbone} through \ref{table:horrebowbackbone}.
Figure \ref{fig:numberofdays} shows the number of days within each year covered by (a) the different BB series
 (i.e. including both primary and secondary observers) and by (b) our final composite series.
One can see that the coverage is very good after ca 1800, but very poor in 1780--1795.
This poorly covered period has led to large uncertainties in the daisy-chain method in the 18th century.

\begin{table*}
	\centering
	\caption{Backbones used in this study: name and identification (Id) of the primary BB observer; period covered by the BB composite series;       period of observations of the primary observer; number of observers included in the BB series;       number $N_{\rm d}$ of daily observations of the BB composite series; and direct daily overlap with the reference BB series, i.e. the number of days available in both BB series $M_{\rm d}$.}
	\label{table:backbones}
	\begin{tabular}{lccccccc}
		\hline
		Backbone & Id & \multicolumn{2}{c}{Period of observations} & Observers & $N_{\rm d}$ & $M_{\rm d}$\tablefootmark{a}\\
		 &  & series & primary observer &  & & \\
		\hline
		\small
		RGO 		& 332 &1882--2010 & 1900--1976  & 81 & 46087 & 	  \\
		Kanzelh\"ohe	& 606 & 1935--2010 & 1957--2010  & 156& 25690 & 25526\\
		Wolfer 		& 335+338&1858--1919 & 1882--1910  & 25 & 22968 & 16601\\
		Schmidt 	& 292 &1841--1900 & 1841--1883  & 11 & 18240 & 11708\\
		Schwabe 	& 279 &1822--1892 & 1825--1867  & 22 & 14160 & 7386 \\
		Flaugergues & 22+227&1774--1844 & 1788--1830  & 12 & 6948  & 1383 (1503)\\
		Horrebow 	& 180&1749--1799 & 1761--1776  & 4  & 2762  & 1775 (1795)\\
		Schubert 	& 178&1754--1758 & 1754--1758  & 1  & 492   & 10   (20)\\
		Zucconi 	& 177&1754--1760 & 1754--1760  & 1  & 899   & 17   (29)\\
		Hagen 		& 161&1739--1751 & 1739--1751  & 1  & 116   & 21   (34)\\
		\hline
	\end{tabular}
	\tablefoot{\\
	\tablefoottext{a}{Values in parenthesis are within $\pm1$ day interval.}}
\end{table*}

\subsection{Construction of the backbone series}
\label{s:mat}
We started by building a direct calibration matrix \citep[cf.][]{usoskin_dependence_2016} between the secondary observer to be calibrated
 and the primary BB observer for the days when both have observations.
If, on a given day, $N_1$ and $N_2$ groups were recorded by the primary and secondary observers, respectively, then unity was added to
 the row $N_1$ and column $N_2$ of the matrix.
In this way, the matrix was filled with all the overlapping days.
Then the matrix was normalized such that each of its values were divided by the total sum over the corresponding column.
Thus, we obtained a matrix of probability density functions (PDF) to find a value of $G^*$ reported by the primary observer for each day with the given value $G$ reported by the secondary observer.
This allows a direct calibration of the secondary observer to the primary observer by replacing the $G$ value with the PDF of $G^*$.
This is the most straightforward method for calibration applied directly to the data.

However, this matrix can potentially have some gaps due to poor statistics and limited range of overlap between the observers.
In such cases, we fill the gaps by fitting the statistically significant part of the matrix with a function
\begin{equation}
\label{eq:exp}
\langle G^*\rangle- G =R_0+Be^{-a G}\, ,
\end{equation}
where $\langle G^*\rangle$ are the mean counts of the primary observer (i.e., the mean of the PDF of each column of the matrix)
 for a given count of the secondary observer $G$, $R_0$, $B$, and $a$ are constants calculated for each pair of observers individually.
We used the weighted least mean squares to find the best-fit parameters.
This functional shape (asymptotic exponential approach to a constant offset in the difference) was proposed by \citet{usoskin_dependence_2016}
 and found suitable for this kind of dependence, using synthetic data that were based on RGO sunspot group area data.

Only those columns of the matrix that contain more that 20 overlapping days were included into the fitting procedure.
If the fit deviated by more than one group from the actual mean $\langle G^*\rangle$,
 such columns were excluded, and the fit was redone.
In such cases we refilled the column matrices using a PDF derived with a bootstrap Monte Carlo (MC, hereafter) simulation.
For this, we randomly selected half of the overlapping days from the two observers, reconstructed the matrix using this half-statistics and
 recalculated the fit for the matrix.
This process was repeated 1000 times.
The result of this simulation was used as a PDF for the corresponding column in the matrix.

\begin{figure}
	\centering
\begin{overpic}[width=1\linewidth]{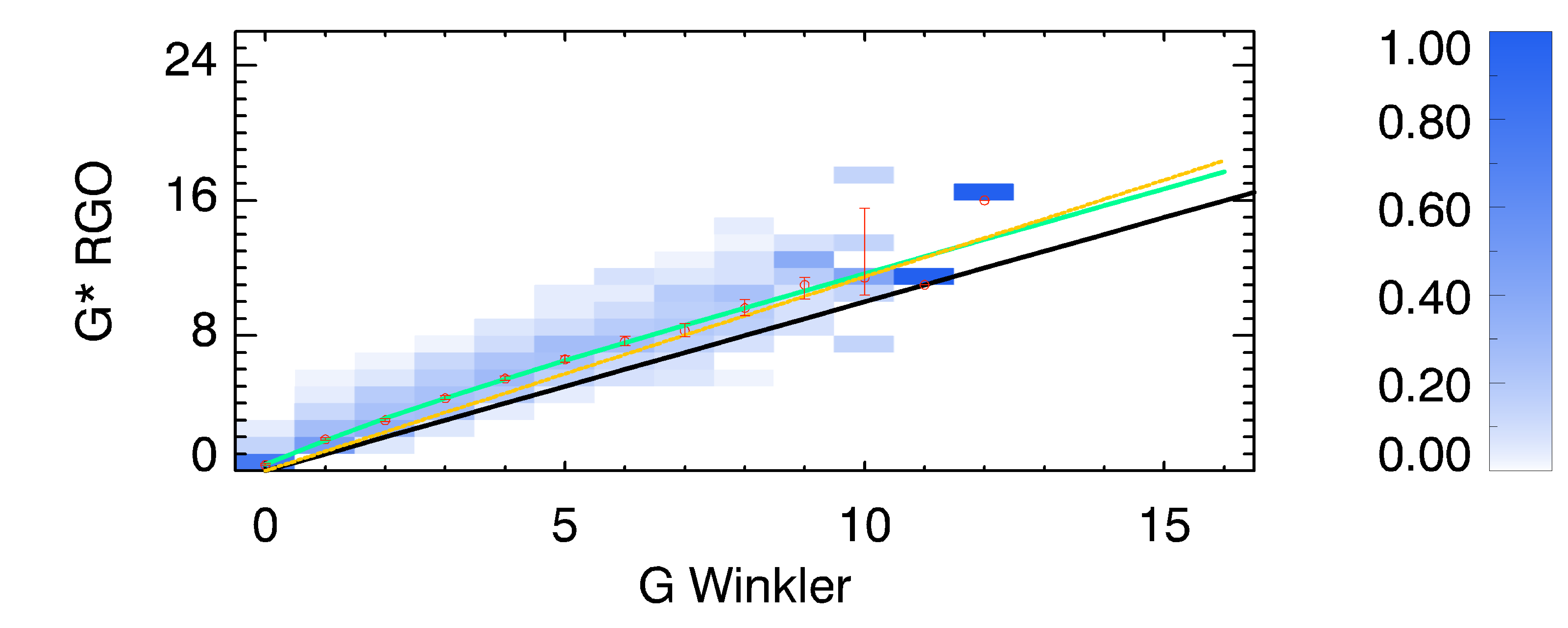}
	\put (17,34) {a)}
\end{overpic}
\vskip 0.5cm
\begin{overpic}[width=1\linewidth]{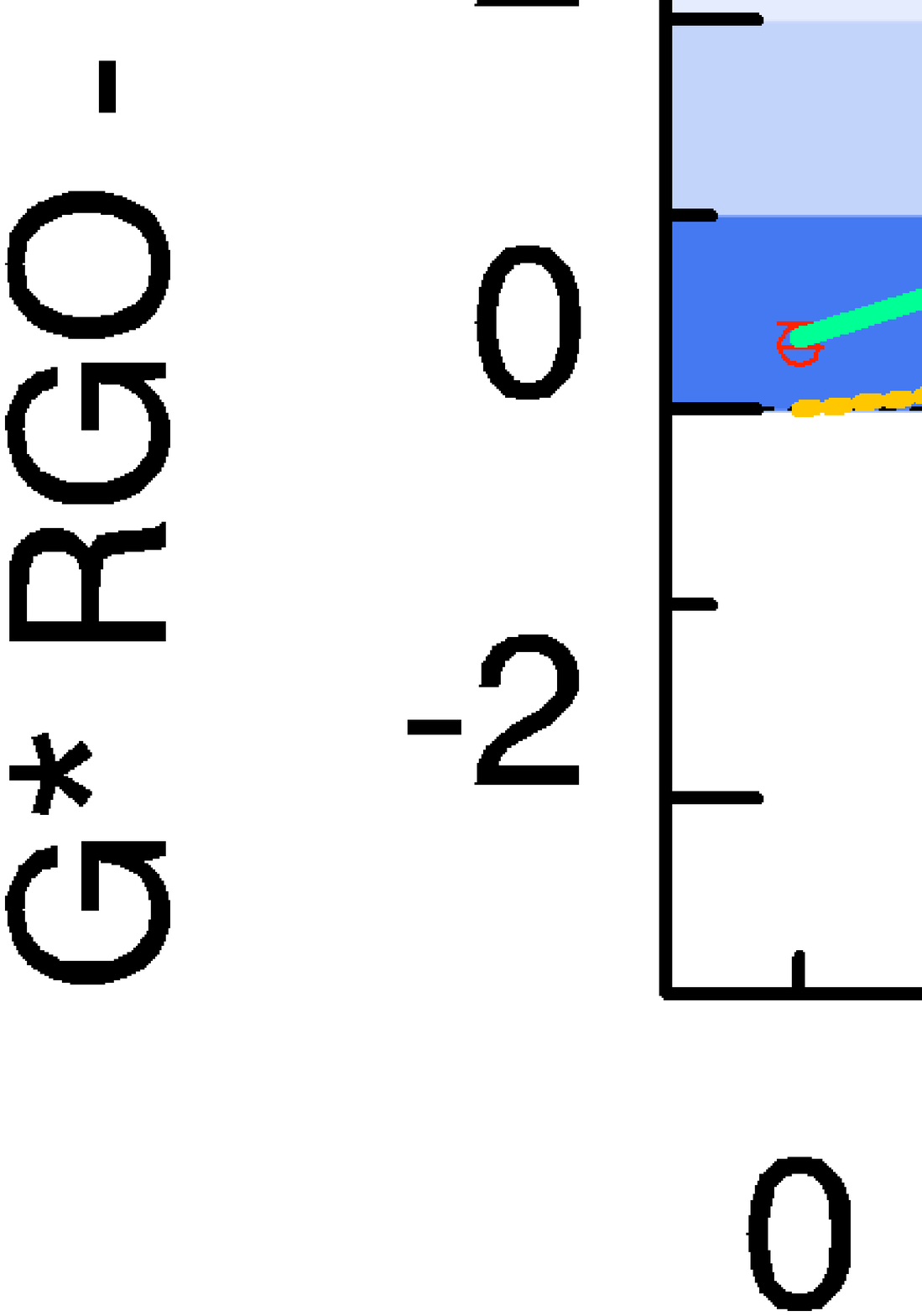}
	\put (17,34) {b)}
\end{overpic}
\vskip 0.5cm
\begin{overpic}[width=1\linewidth]{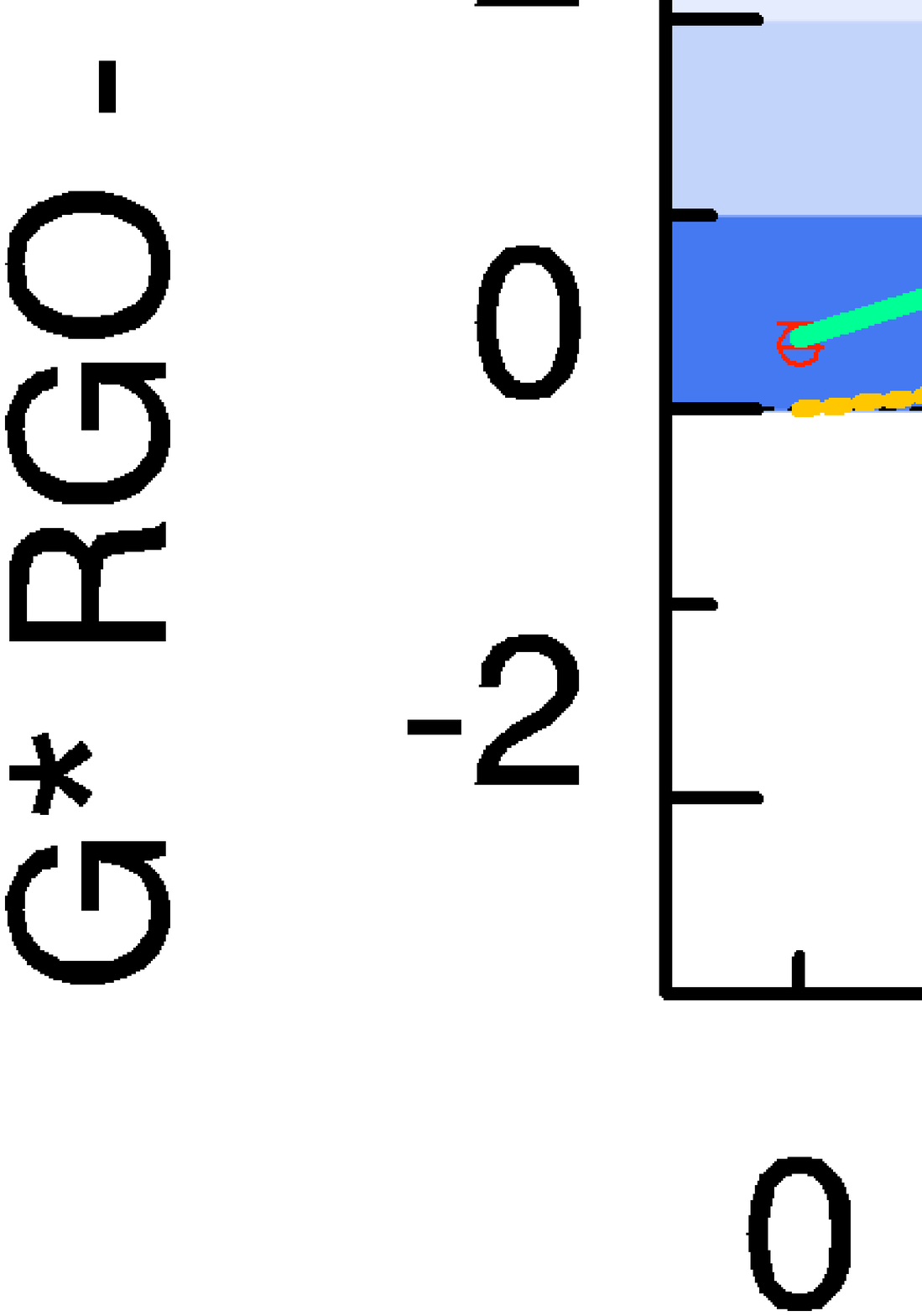}
	\put (17,34) {c)}
\end{overpic}
	\caption{Example of the construction of the calibration matrix for Winkler (secondary observer, $G$) to RGO (primary, $G^*$) over 1900--1910. Panel
(a) shows the original distribution matrix $G^*$ vs. $G$: the black line has a slope of unity. Panel
(b) shows the difference, $G^*-G$ vs. $G$. Panel
(c) is the same as (b) but the empty columns for $G^*>8$ have been filled with the results of the MC simulation.
The red circles with error bars depict the mean $G^*$ values for each $G$ column and their 1$\sigma$ uncertainty.
The yellow line shows the $k-$factor used in \citet{hoyt_group_1998}.}
	\label{fig:WOLFERrgo}
\end{figure}

An example of the matrix is shown in Figure~\ref{fig:WOLFERrgo}a for Winkler (secondary
 observer, $G$) and RGO (primary reference observer, $G^*$) over the period of their overlap (1900 -- 1910 with 2480 common days).
It is apparent that RGO typically reported more groups than Winkler for the same day, since most of the matrix values lie above the
 line expected for a perfect match between the two
 (black line). 
The matrix of the difference, $G^*-G$ versus $G$, is shown in Fig. \ref{fig:WOLFERrgo}b.
The red circles with error bars represent the mean $\langle G^*\rangle$ value in each
 $G$ column and its (asymmetric) $1\sigma$ intervals.
The green curve shows the best fit of the functional form of Eq.~\ref{eq:exp}.
It is obvious that the relation between $G^*$ and $G$ is non-linear and cannot be represented
 by a simple linear scaling $k-$factor.
One can see that, because of the limited overlap, the matrix is well constructed only for $G<9$.
For higher values, the fit (Eq.~\ref{eq:exp}) has to be used.
The full matrix with the values filled with the MC method for $G>8$ is shown
 in Fig. \ref{fig:WOLFERrgo}c.

Each secondary observer was calibrated to the BB observer by replacing, from the matrix, every daily count $G$ with the PDF of the
  calibrated counts $G^*$.
In this way we directly convert the observations of the secondary observer to the BB condition
 without making any assumption about the type of relationship (e.g. linearity) and with a straightforward error estimate.

For each BB we constructed a composite series by averaging all the PDFs of all the available observations for every day, so that
 again, instead of one count for each day, we get a distribution based on all available observers.
This composite of averaged PDFs includes possible errors in a straightforward way.

Only observers with a sufficiently long record of relatively good quality were included into the analysis.
The selection of secondary observers was made using the following criteria:
\begin{enumerate}
\item
The overlap with the primary BB observer should be not less than 20 common days of observations.
This criterion was not applied for early years (see Section~\ref{sec:Sch}).
\item
Observers with an overall record longer than 10 years were considered only if their overlap with the primary BB observer was at least 4 years.
This is merely to make sure that long-running observers are not calibrated with a small fraction of their observations that might not be representative.
\item
In cases in which we need to perform the fit to extrapolate to missing values in the matrix, we requested the conversion matrix for
 a selected observer to have sufficient data to cover at least three $G-$value bins.
This is necessary since the function described by Eq.~(\ref{eq:exp}) has three parameters.
\item
The matrix should cover, with sufficient statistics, at least one-quarter of the range of counts reported by the secondary observer.
\item
Observers were excluded from the analysis if the difference matrix (see an example in Fig.~\ref{fig:WOLFERrgo}b)
 had an average offset of more than two groups for the $G$ values from 0 to 5.
\item
Observers, whose data could not be fitted accurately enough ($\chi^2$ per degree of freedom $<6$), were also excluded.
\end{enumerate}

After the calibration process of all observers, we compared each individual observer with the composite BB series they were part of. We excluded those that showed significant
 and systematic discrepancies.
Four observers were removed as they showed such differences, namely Taipei observatory (Id 456), Lunping (Id 457), Mojica, Cochabamba, Bolivia (Id 628),
 and XE (Id 715).
We also excluded the Locarno station (Id 614), because of the possible lack of stability after 1980 \citep{clette_revised_2016}.

There are also some special cases, which are described below in detail.

\subsubsection{Sparse data: Schwabe and earlier backbones}
\label{sec:Sch}
Because of the lack of data for the first years of the Schwabe BB, we have not applied the criterion 1 from the list above
 to his data.
Furthermore, while constructing the calibration matrix we considered observations not only during overlapping days but also within $\pm$1
 day; if there was no direct overlap, we first checked one day earlier and then one day later, making sure that no more than one pair entered
 the matrix.
Possible errors due to short-lived groups are negligible compared to the gain of the increased statistical sample
 \citep{willis_re-examination_2016,usoskin_new_2016}.
These relieved constraints were also applied to the BBs covering earlier periods, when the statistics were poor.

\subsubsection{Correcting for low quality observations: Flaugergues, Schubert, Zucconi, and Hagen backbones}

For most BBs, we were able to match observers with a relatively similar quality.
This was not the case for Flaugergues, though.
Flaugergues' data are very important, because they are the only record covering a relatively extended period in the early 1800s.
However, the $G$ values he reported are significantly lower than those by other observers during that period, implying that his observations are of lower quality (higher acuity observational threshold).
Therefore, a calibration of all other observers, with higher quality data, directly to Flaugergues would reduce
 their quality while increasing the uncertainties.
In order to avoid that, we made use of a corrected Flaugergues series, calibrated to the mean level of the other observers of the period.
In order to make the correction, we assumed that the acuity threshold for Flaugergues is $A=100$ msd, which is greater than for any other observer
 \citep{usoskin_new_2016}.
In this case the acuity threshold for Flaugergues does not even have to be the correct one, but it only should allow us to calibrate the overlapping
 observers without downgrading their quality.
Applying the 100 msd threshold and the method described in \citet{usoskin_dependence_2016}, we obtained the following parameters for
 Eq.~(\ref{eq:exp}) for Flaugergues: $a= 0.18$, $R=6.94$, and $B=6.03$.
Then other observers were calibrated to this `corrected' Flaugergues series.

The same process with the same threshold was used for the Schubert, Zucconi, and Hagen BBs.

\subsection{Inter-calibration of backbone series}
\label{sec:bbcalibration}
Once the BB series were constructed and calibrated to the primary BB observer, different BB
 series had to be inter-calibrated to each other.
We used the RGO BB as the reference one, and the others were calibrated to it using a daisy chain.
The calibration of the BB series was performed using a procedure similar to that for the individual observers,
 by constructing the cross-calibration matrix between the whole BB series this time.
However in this case, we have, for each day, not a single $G$ value but a PDF from each observer (now the entire composite BB
  series is considered an observer).
In order to account for that, we constructed the calibration matrix using a MC simulation as described below.
For each day with simultaneous observations from both `observers' (the BB series), we randomly selected $G$ values corresponding
 to the PDFs and filled the matrix.
This process was performed 1000 times for each day, and the final matrix was computed as the average among all the individual matrices.

Monte Carlo simulations were used to calibrate the secondary BB to the reference one accounting for the error propagation.
We randomly picked a $G$ value from the PDF for each day of the secondary BB series and obtained, from the matrix,
 the PDF of the $G^*$ values for the reference BB.
This was repeated 1000 times and the average PDF of the $G^*$ values was considered as the calibrated PDF of the
 secondary BB series for that day.

\begin{figure}
	\centering
	\begin{overpic}[width=1\linewidth]{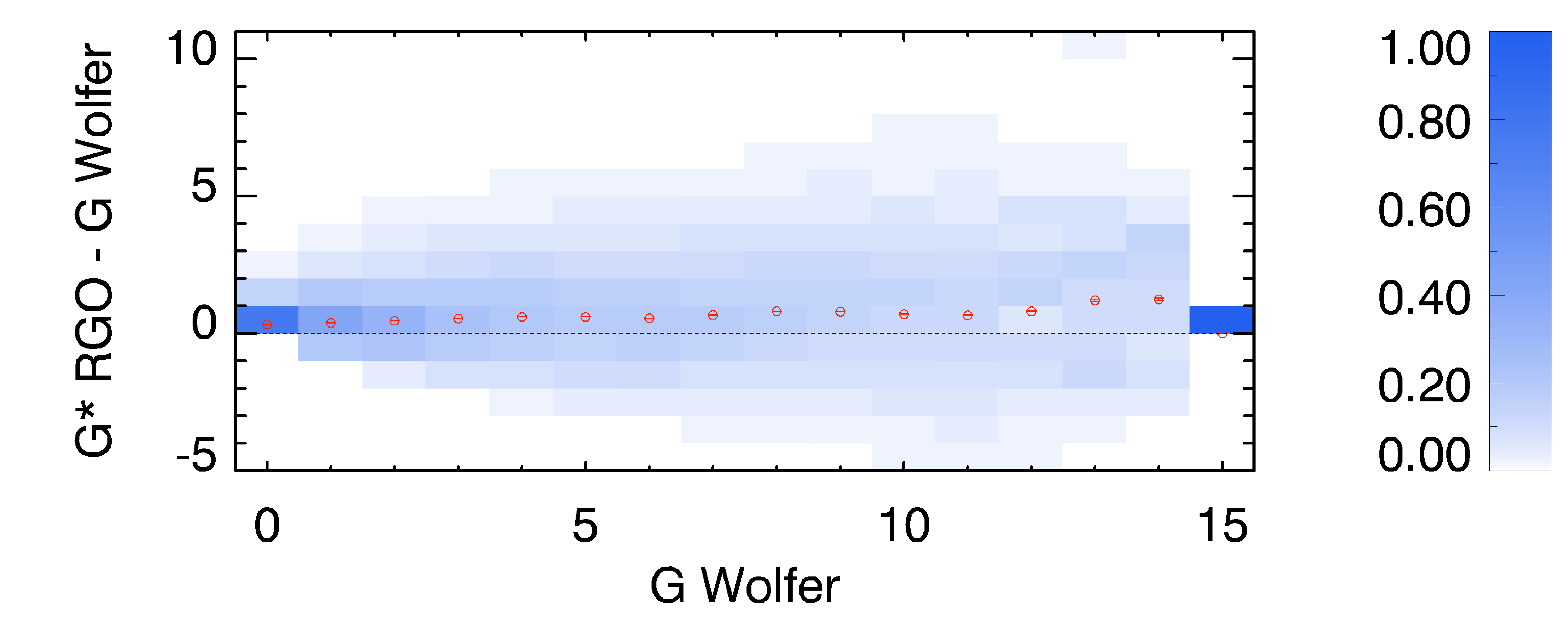}
		\put (17,34) {a)}
	\end{overpic}
\vskip 0.5cm
	\begin{overpic}[width=1\linewidth]{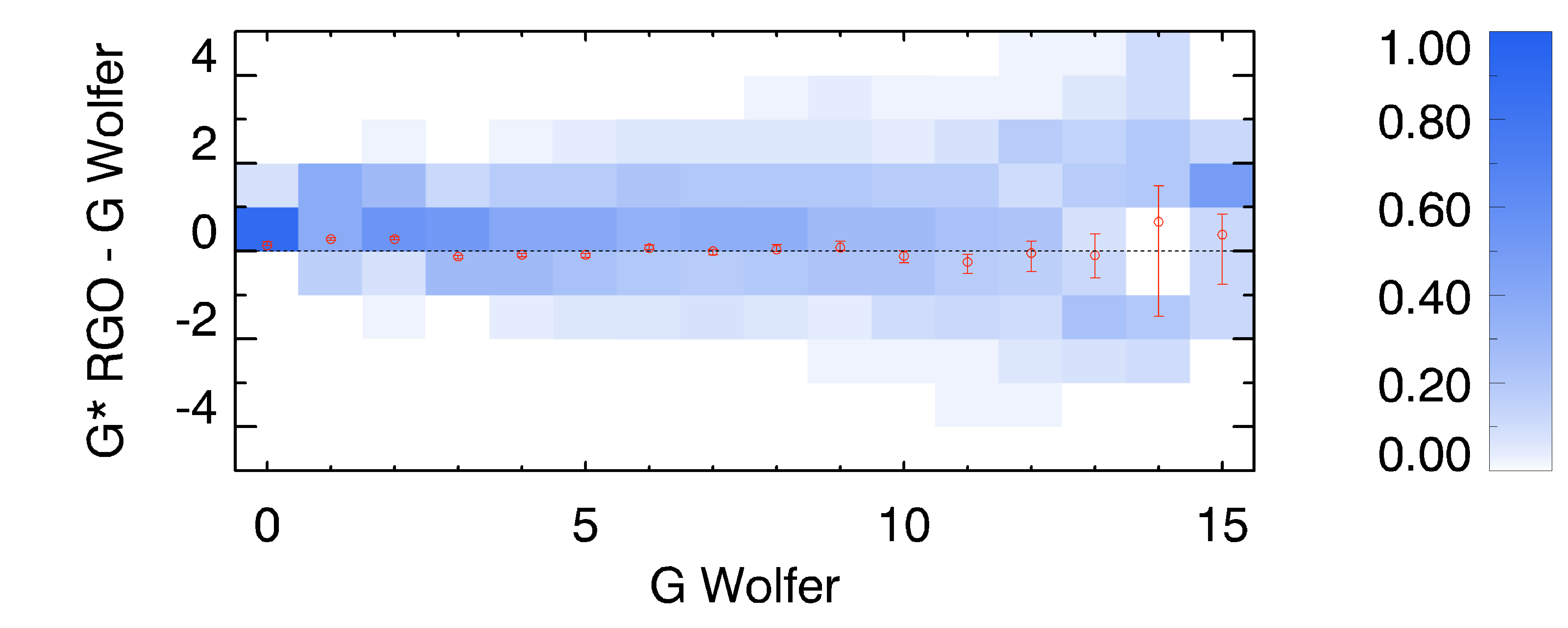}
		\put (17,34) {b)}
	\end{overpic}
	\caption{Difference between the Wolfer and RGO backbones.
Panel (a) shows an uncalibrated matrix after the full MC filling;
   panel (b) shows the same matrix after the calibration.
    The red circles depict the average values in every column with their 1$\sigma$ uncertainty ranges.}
	\label{fig:backbone_wolferrgo}
\end{figure}

The procedure is illustrated in Fig.~\ref{fig:backbone_wolferrgo}, which shows the result of
 the calibration of the secondary Wolfer BB series to the primary RGO BB series.
It is evident from the panel a) that the RGO BB $G$ values are systematically higher than those
 of the Wolfer BB (the difference is positive), implying that RGO is a better observer than Wolfer.
After the calibration (panel b), the two series match each other so that the mean difference is
 consistent with zero in the entire range of $G$ values implying that the calibration was carried out correctly.

This procedure works well for all the BBs. However, the results for the Horrebow BB series are very uncertain. 
The overlap of this series with the Flaugergues BB series is short and occurs only during activity minima around 1775 and 1795, which
 gives merely four points ($G$ values) to perform  the fit and to extrapolate to the rest of the range of values.
Since the method gives a realistic estimate of the uncertainties, this is clearly expressed in
 large error bars for the 18th century.

\subsection{Construction of the final series}

After all the BBs were calibrated to the reference RGO series,
 the final composite series was produced.
First, for each day, all the available BB series values (in the form of a PDF) were merged into a single
 PDF for that day.
From the daily PDFs of the calibrated $G$ values we produced the monthly $G$ values using a MC simulation.
For this, for each day with available data within a month, we randomly selected a $G$ value from the
 final daily PDF and then computed the monthly value as the arithmetic mean of these daily values.
This procedure was repeated 1000 times, and the PDF of the monthly values was constructed for each month.
This MC method considers all the uncertainties straightforwardly.
Finally, we collected the mean and asymmetric $\pm 1\sigma$ uncertainty level (a Table is available at the CDS).

\begin{figure*}
	\centering
	\vskip 0.5cm
	\includegraphics[width=1\linewidth]{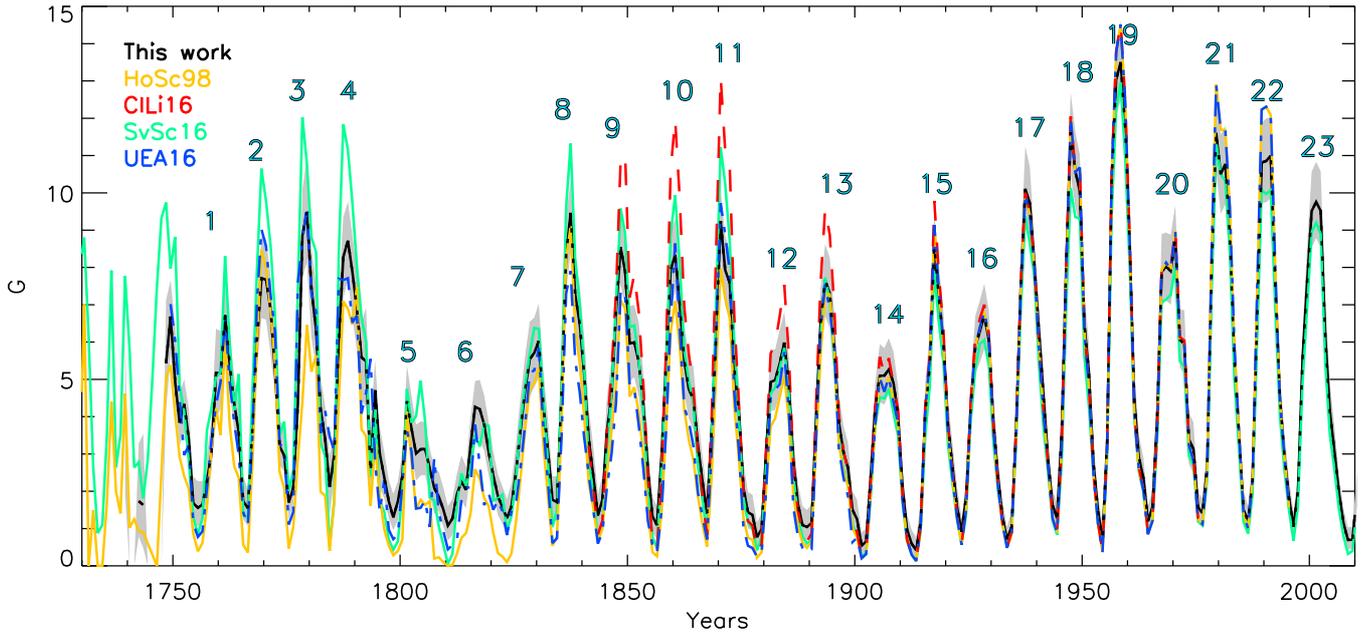}
	\caption{Annually averaged number of sunspot groups. This work is indicated in black with the $\pm 1\sigma$ area shaded; HoSc98 is indicated in yellow;
		UEA16 is shown in blue; SvSc16 is shown in green; and ClLi16 is indicated in red.
		Numbers on top of the curves denote the conventional solar cycle numbering.}
	\label{fig:GSNseries}
\end{figure*}

Next, the annual numbers of sunspot groups with their asymmetric $\pm 1\sigma$ uncertainties were calculated from the monthly
 values in the same manner as monthly values from the daily values.
The final annual series is given in Table~\ref{table:series} and shown in Figure~\ref{fig:GSNseries}.
The GSN in years without reliable values are denoted by -99.
\begin{figure*}
	\centering
	\vskip 0.5cm
	\includegraphics[width=1\linewidth]{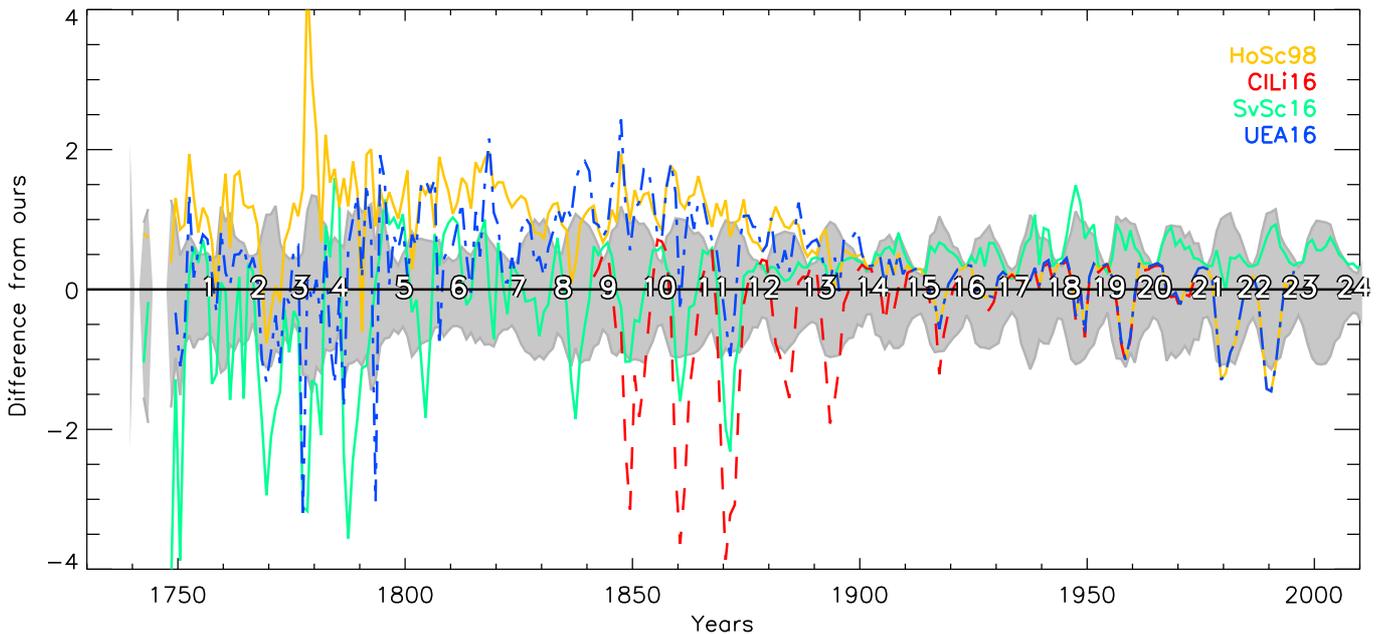}
	\caption{Differences of the annual GSN between our
		series and other series (as denoted in the legend). Positive values imply that our series is higher.
		The grey shading denotes the $\pm 1\sigma$ range of our series.
		The numbers denote the conventional solar cycle numbering.}
	\label{fig:seriescomparison_y_difference}
\end{figure*}

\begin{table*}
	\centering
	\caption{Annual values of the proposed GSN series with the asymmetric $1\sigma$ intervals.}
	\label{table:series}
	\begin{tabular}{|cccc|cccc|cccc|cccc|}
		\hline
		Year & G & $\sigma_+$&$\sigma_-$ & Year & G & $\sigma_+$&$\sigma_-$& Year & G & $\sigma_+$&$\sigma_-$& Year & G & $\sigma_+$&$\sigma_-$\\
		\hline
1739& 4.01&2.29&2.07&1799& 1.74&0.52&0.49&1859& 7.94&0.94&1.02&1919& 6.00&0.77&0.79\\
1740& -99&-99&-99&1800& 2.41&0.66&0.60&1860& 8.34&0.95&0.99&1920& 3.78&0.70&0.65\\
1741& -99&-99&-99&1801& 4.42&0.93&0.84&1861& 7.01&0.97&0.96&1921& 2.65&0.65&0.58\\
1742& 1.73&1.34&1.00&1802& 3.69&0.88&0.72&1862& 5.50&0.91&0.88&1922& 1.59&0.51&0.40\\
1743& 1.63&1.92&1.13&1803& 3.01&0.75&0.70&1863& 4.74&0.87&0.86&1923& 0.92&0.38&0.32\\
1744& -99&-99&-99&1804& 3.13&0.75&0.72&1864& 4.47&0.88&0.85&1924& 1.85&0.51&0.46\\
1745& -99&-99&-99&1805& 3.13&0.71&0.68&1865& 3.23&0.87&0.76&1925& 4.22&0.76&0.70\\
1746& -99&-99&-99&1806& 2.62&0.58&0.68&1866& 2.27&0.71&0.58&1926& 5.87&0.87&0.80\\
1747& -99&-99&-99&1807& 2.09&0.47&0.67&1867& 1.41&0.54&0.46&1927& 6.28&0.76&0.80\\
1748& 5.43&1.66&1.27&1808& 1.86&0.65&0.43&1868& 3.62&0.86&0.71&1928& 6.72&0.88&0.92\\
1749& 6.68&0.95&0.98&1809& 1.52&0.54&0.44&1869& 6.21&0.99&0.88&1929& 6.05&0.80&0.73\\
1750& 4.94&1.53&0.44&1810& 1.08&0.49&0.42&1870& 9.24&0.88&1.02&1930& 3.83&0.77&0.64\\
1751& 3.84&0.77&0.62&1811& 1.27&0.55&0.42&1871& 7.93&0.89&0.89&1931& 2.39&0.53&0.48\\
1752& 4.34&0.68&0.75&1812& 1.92&0.51&0.61&1872& 7.58&0.85&0.93&1932& 1.31&0.42&0.34\\
1753& 3.40&0.80&0.70&1813& 2.26&0.70&0.49&1873& 5.27&0.95&0.79&1933& 0.72&0.35&0.27\\
1754& 1.68&0.72&0.52&1814& 2.04&0.62&0.51&1874& 4.18&0.78&0.77&1934& 1.05&0.40&0.32\\
1755& 1.56&0.70&0.51&1815& 3.22&0.76&0.64&1875& 2.09&0.66&0.51&1935& 3.75&0.75&0.64\\
1756& 1.64&0.62&0.48&1816& 4.26&0.75&0.78&1876& 1.44&0.58&0.43&1936& 7.45&0.91&0.80\\
1757& 2.28&0.68&0.48&1817& 4.20&0.80&0.76&1877& 1.37&0.55&0.42&1937& 10.10&1.15&1.03\\
1758& 3.02&0.93&0.49&1818& 3.78&0.80&0.75&1878& 0.77&0.44&0.30&1938& 9.72&0.96&1.06\\
1759& 5.17&1.17&1.19&1819& 3.04&0.76&0.62&1879& 0.99&0.42&0.35&1939& 8.10&0.76&0.76\\
1760& 5.17&1.07&0.97&1820& 2.42&0.63&0.53&1880& 3.03&0.75&0.64&1940& 6.31&0.72&0.76\\
1761& 6.72&0.74&1.16&1821& 1.87&0.58&0.50&1881& 4.93&0.89&0.81&1941& 4.55&0.74&0.67\\
1762& 5.44&0.87&0.73&1822& 1.56&0.60&0.42&1882& 4.98&0.81&0.77&1942& 2.86&0.63&0.50\\
1763& 4.35&0.77&0.69&1823& 1.28&0.55&0.36&1883& 5.43&0.95&0.78&1943& 1.63&0.43&0.35\\
1764& 3.58&0.72&0.70&1824& 1.60&0.64&0.39&1884& 5.98&0.87&0.79&1944& 1.27&0.43&0.33\\
1765& 1.73&0.67&0.42&1825& 2.54&0.72&0.60&1885& 4.88&0.74&0.80&1945& 3.55&0.69&0.61\\
1766& 1.55&0.48&0.46&1826& 3.67&0.93&0.77&1886& 2.79&0.70&0.63&1946& 8.07&0.94&0.86\\
1767& 3.64&0.71&0.56&1827& 4.71&0.96&0.85&1887& 1.66&0.54&0.49&1947& 11.62&1.10&1.17\\
1768& 6.02&0.95&0.81&1828& 5.54&0.97&0.95&1888& 1.10&0.49&0.34&1948& 10.59&0.98&1.00\\
1769& 7.71&1.16&0.99&1829& 5.71&0.95&0.94&1889& 1.04&0.47&0.39&1949& 10.04&1.01&0.91\\
1770& 7.68&1.14&1.00&1830& 6.03&0.99&0.99&1890& 1.15&0.47&0.39&1950& 6.47&0.88&0.86\\
1771& 6.89&0.97&1.14&1831& 4.34&0.94&0.85&1891& 3.98&0.73&0.71&1951& 5.19&0.79&0.76\\
1772& 5.23&0.96&0.66&1832& 3.08&0.81&0.69&1892& 6.51&0.95&0.84&1952& 2.74&0.55&0.55\\
1773& 3.21&0.70&0.50&1833& 1.77&0.61&0.53&1893& 7.66&0.94&1.03&1953& 1.46&0.48&0.42\\
1774& 2.96&0.79&0.41&1834& 1.70&0.71&0.50&1894& 7.31&0.95&0.90&1954& 0.74&0.32&0.27\\
1775& 1.70&0.49&0.47&1835& 4.69&0.91&0.82&1895& 5.86&0.98&0.81&1955& 3.33&0.65&0.55\\
1776& 2.08&0.61&0.42&1836& 8.32&1.08&0.95&1896& 3.85&0.76&0.68&1956& 10.29&1.03&1.05\\
1777& 4.33&1.20&0.60&1837& 9.47&0.83&1.17&1897& 3.05&0.70&0.62&1957& 13.03&1.01&0.97\\
1778& 8.86&1.19&1.13&1838& 7.29&1.00&1.05&1898& 2.63&0.70&0.59&1958& 13.50&1.05&1.10\\
1779& 9.48&1.54&1.40&1839& 6.49&1.03&0.93&1899& 1.50&0.52&0.43&1959& 11.71&0.91&0.93\\
1780& 7.45&1.06&1.32&1840& 5.12&1.01&0.86&1900& 1.25&0.50&0.41&1960& 8.53&1.05&0.96\\
1781& 6.33&1.15&1.00&1841& 3.40&0.77&0.77&1901& 0.54&0.33&0.25&1961& 4.45&0.75&0.75\\
1782& 4.20&0.82&0.95&1842& 2.42&0.79&0.60&1902& 0.65&0.34&0.28&1962& 2.91&0.57&0.57\\
1783& 3.41&0.89&0.63&1843& 1.35&0.62&0.42&1903& 2.36&0.66&0.46&1963& 2.35&0.53&0.45\\
1784& 2.12&0.83&0.76&1844& 1.78&0.68&0.54&1904& 4.23&0.67&0.66&1964& 1.20&0.42&0.30\\
1785& 2.97&0.54&0.67&1845& 3.61&0.86&0.80&1905& 5.10&0.84&0.70&1965& 1.58&0.46&0.38\\
1786& 6.00&1.29&0.59&1846& 4.57&0.94&0.85&1906& 5.10&0.77&0.81&1966& 3.97&0.68&0.66\\
1787& 8.28&1.12&1.07&1847& 6.82&0.92&1.18&1907& 5.27&0.83&0.70&1967& 7.88&0.98&0.90\\
1788& 8.72&1.01&0.99&1848& 8.55&0.87&0.97&1908& 4.91&0.79&0.77&1968& 8.03&0.93&0.90\\
1789& 7.87&1.03&1.31&1849& 7.89&1.06&0.90&1909& 4.06&0.79&0.64&1969& 7.90&0.98&0.89\\
1790& 6.88&1.03&1.08&1850& 5.96&0.93&0.91&1910& 2.12&0.53&0.48&1970& 8.75&0.89&0.85\\
1791& 5.59&0.96&1.14&1851& 5.99&0.98&0.96&1911& 0.97&0.43&0.35&1971& 6.07&0.77&0.81\\
1792& 5.48&1.32&1.18&1852& 5.48&1.03&0.96&1912& 0.60&0.36&0.24&1972& 5.94&0.88&0.85\\
1793& 2.51&1.34&0.52&1853& 4.33&0.87&0.82&1913& 0.42&0.35&0.20&1973& 3.40&0.66&0.68\\
1794& 4.71&0.81&1.20&1854& 2.55&0.82&0.66&1914& 1.17&0.50&0.34&1974& 3.12&0.68&0.63\\
1795& 3.02&0.69&0.93&1855& 1.33&0.49&0.47&1915& 4.13&0.78&0.74&1975& 1.57&0.43&0.41\\
1796& 2.38&0.73&0.59&1856& 1.10&0.60&0.38&1916& 5.36&0.88&0.86&1976& 1.41&0.39&0.35\\
1797& 1.73&0.57&0.44&1857& 2.95&0.76&0.70&1917& 8.57&0.79&0.98&1977& 2.65&0.57&0.53\\
1798& 1.30&0.67&0.35&1858& 5.44&1.05&0.87&1918& 7.20&0.95&0.92&1978& 7.92&1.04&0.80\\
\hline
\end{tabular}
\end{table*}

\section{Validation of the results}
\label{s:results}

\subsection{Comparison with other series}

Other published GSN series are also shown in Fig.~\ref{fig:GSNseries}, but without the uncertainties.
While all the series are dominated by the 11-year solar cycle, the centennial variability
 differs among different reconstructions.
The ClLi16 and SvSc16 series are systematically higher than our reconstruction in the 19th and 18th centuries, while the HoSc98 series is somewhat lower.
The present result is close to UEA16 and lies between the `high' and `low' models.

\begin{figure}
	\centering
	\vskip 0.5cm
	\includegraphics[width=1\linewidth]{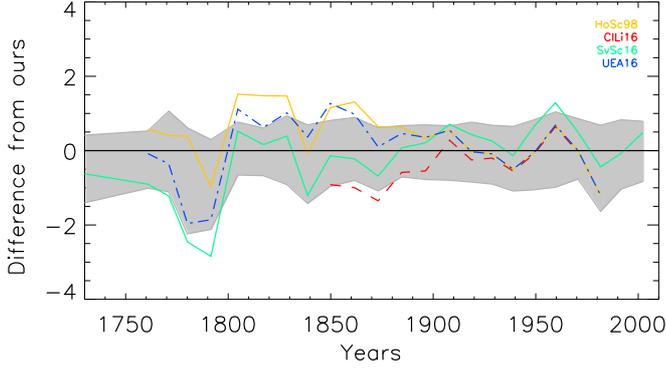}
	\caption{Differences of the solar cycle averaged GSN between our
		series and other series (as denoted in the legend). Positive values imply that our series is higher.
		The grey shading denotes the $\pm 1\sigma$ range of our series.}
	\label{fig:seriescomparison_sc_difference}
\end{figure}

Figures~\ref{fig:seriescomparison_y_difference} \& \ref{fig:seriescomparison_sc_difference} show the
 difference between various other series and the result presented here.

One can see that all the series agree with each other in the 20th century, except the SvSc16 series
 which is systematically lower than all others, although still within the error bars.

The UEA16 series is very close to our series during cycle maxima, while there are noticeable differences around the minima.
The two series diverge for cycles 2 (our series is lower than UEA16), 8-9 (ours is higher),
 and 21-22 (ours is lower).
The differences in cycles 22-23 can be explained by different observers used:
 while UEA16 used only RGO and Koyama over that period, we used here more than 150 observers, which allows
 us to estimate the activity more accurately.

During the solar cycle minima our series agrees with SvSc16, but there are distinct differences during the maxima.
The SvSc16 series gives higher values over the cycles 1-5 and 8-11, while lower values are found for almost all cycles over the 20th century.
These differences can be at least partly explained by the -7\% ad hoc adjustment applied by SvSc16 to the data after 1940 and
 by the choice of Koyama as the reference observer (see also a discussion about this in Sec. \ref{s:discussion}).

Over the 20th century, the ClLi16 series is essentially the same as that of HoSc98, but they deviate over the 19th century so that maxima in the ClLi16 series
 are 3-4 groups higher than in HoSc98, and hence also than in ours.
Keeping in mind that we ignored the RGO data before 1900 and used Wolfer as the reference for that period, the higher values by ClLi16
 suggest a possible overcorrection of the RGO series by these authors. This is in agreement with the findings of \citet{lockwood_assessment_2016}.

\begin{figure}
	\centering
	\includegraphics[width=1\linewidth,bb=67 488 505 737, clip=true]{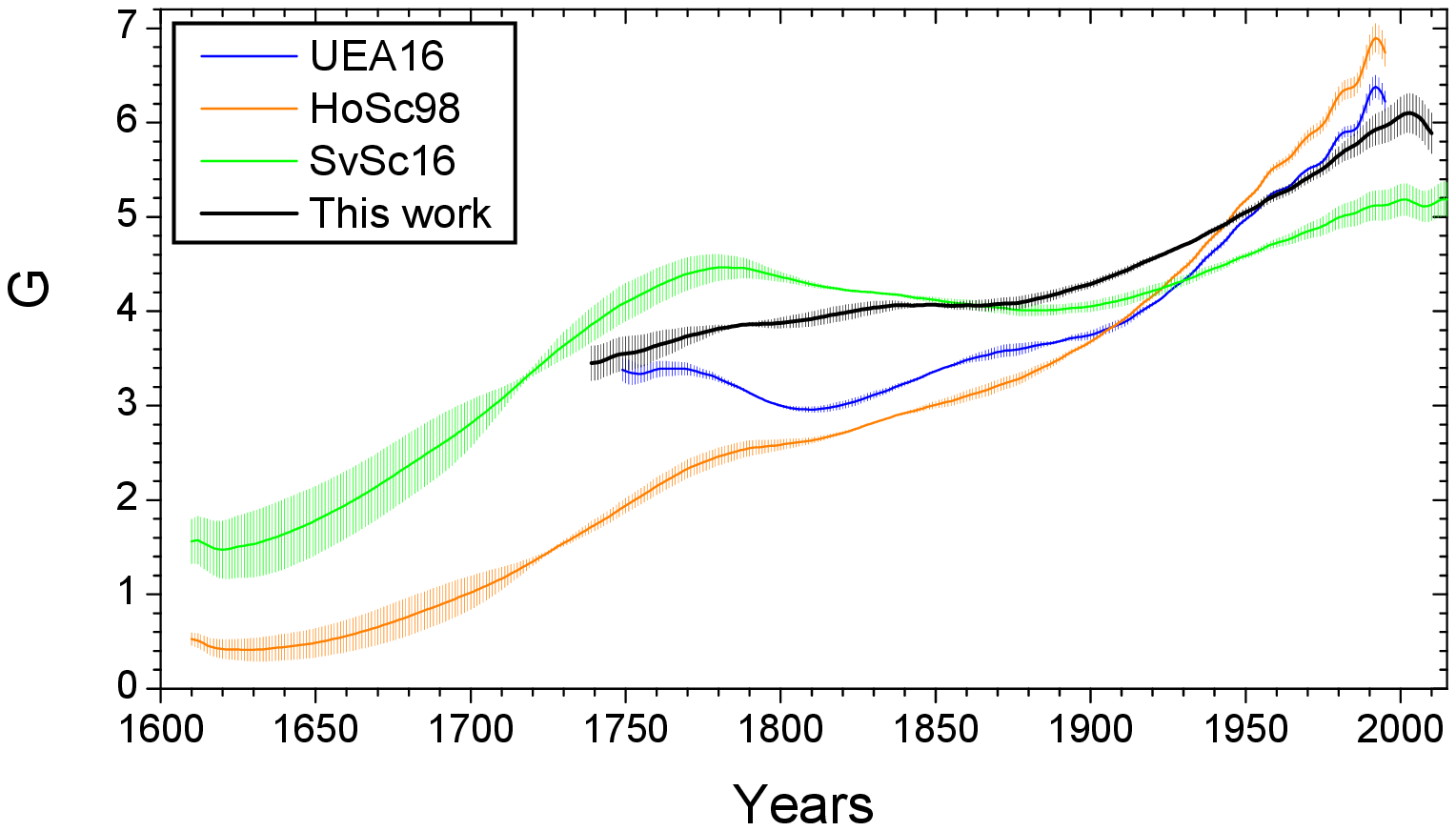}
	\caption{The long-term secular trend in different SN series, studied here, defined as the first SSA component.
		The shading represents only statistical uncertainties of the SSA method.}
	\label{Fig:SSA}
\end{figure}

In Figure~\ref{Fig:SSA} we show the secular trends of different series considered here,
 using the non-parametric SSA \citep[singular spectrum analysis, ][]{vautard_singular-spectrum_1992}.
The SSA method is based on decomposition of a time series into several components with distinct temporal behaviours.
It is very convenient for the identification of long-term trends and quasi-periodic oscillations, especially
 in the conditions when the secular trend is subdominant with respect to the main periodicity.
As the secular trend we consider the first SSA components of the SN series.
We used the time window for the SSA in the range of 80--100 years, where the result is stable.
All series show that the activity level was highest in the late 20th century,
 corresponding to the modern grand maximum, but the relative enhancement differs among series.
The greatest increase over the last 200 years (defined as the ratio of the values in 2000 and in 1750)
  is observed for the HoSc98 series ($\approx 2.6$), followed by the UEA (1.9) and our final series (1.7). Finally, SvSc16 series yields 1.3.
Thus, the modern grand maximum is observed in all series. According to this work, this grand maximum is weaker than that
 in the HoSc16 series but greater than in the SvSc16 series.

\subsection{Tests of stability}
\label{s:discussion}

\subsubsection{Choice of backbone observers}

\begin{figure}
\vskip 0.5cm
	\centering
\begin{overpic}[width=1\linewidth]{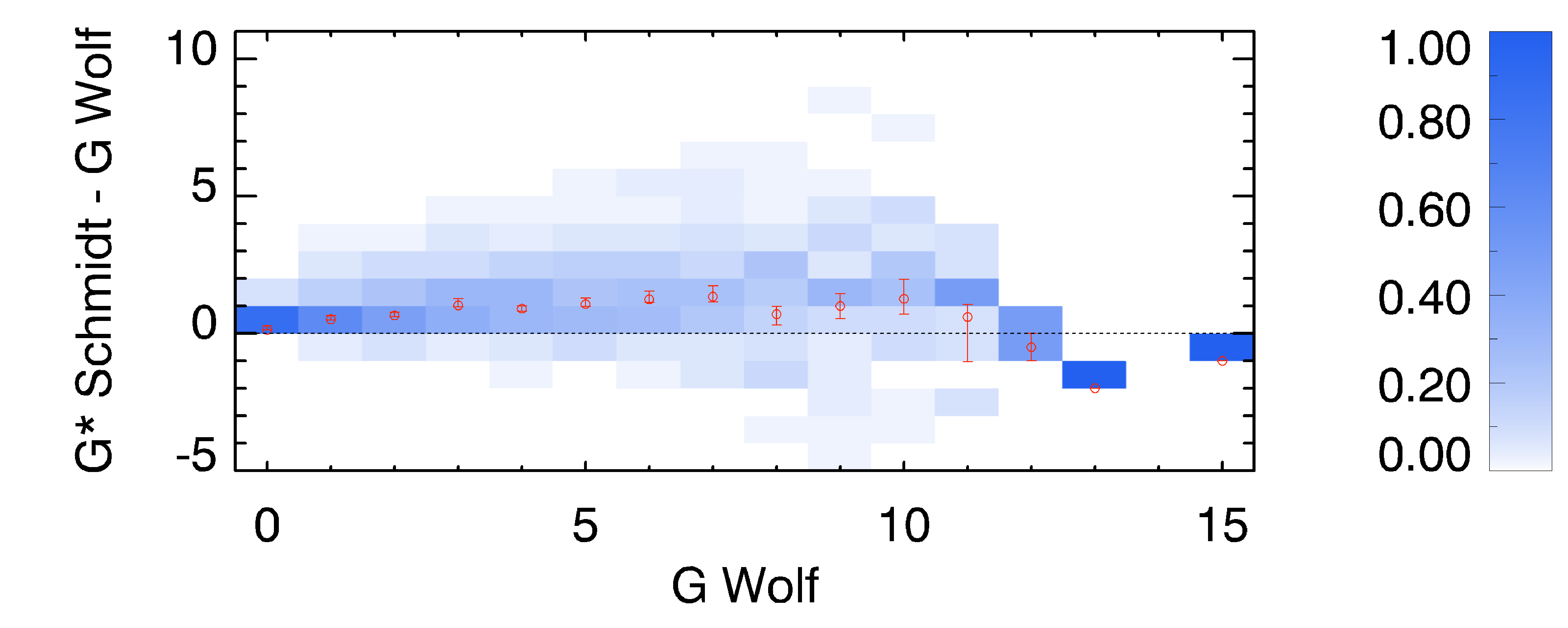}
	\put (17,34) {a)}
\end{overpic}
\vskip 0.5cm
\begin{overpic}[width=1\linewidth]{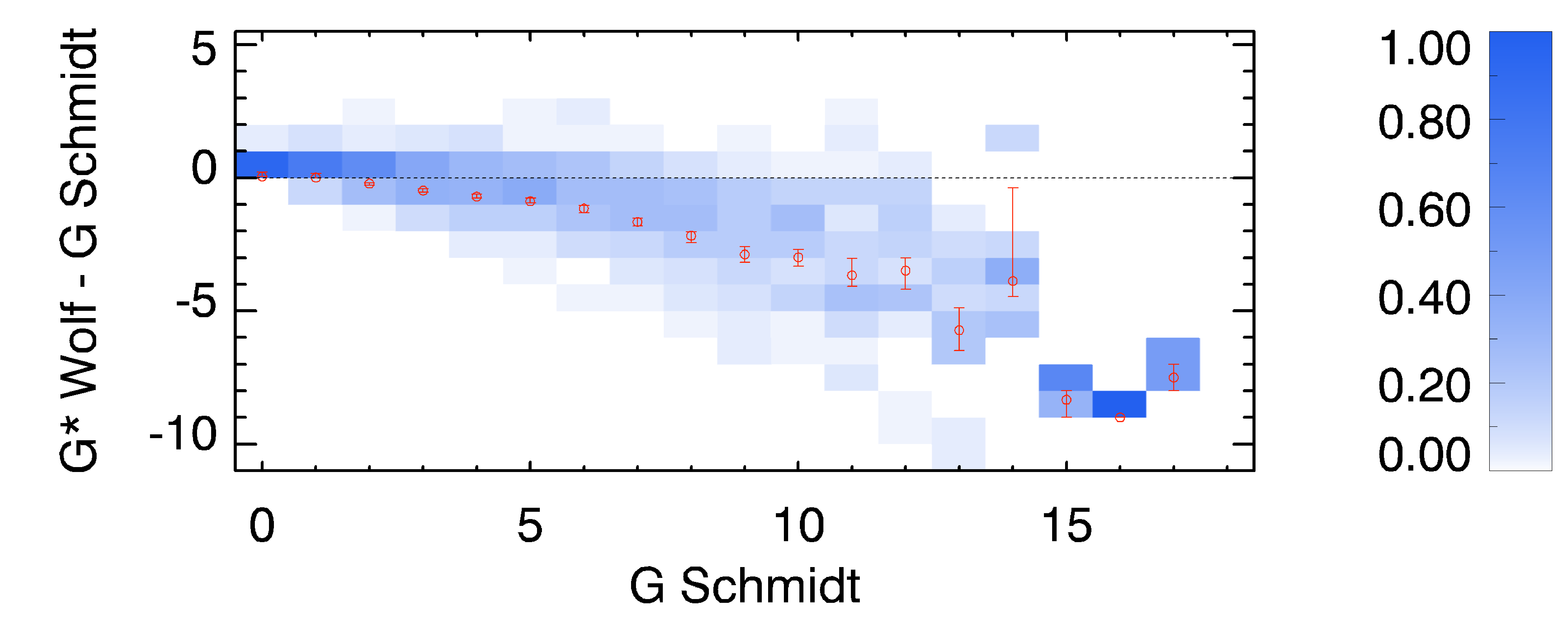}
	\put (17,34) {b)}
\end{overpic}
	\caption{Matrices of the $G$ value difference between Wolf and Schmidt, where Schmidt (panel a) and Wolf (panel b) are selected as  reference observers.}
	\label{fig:schmidtwolfercomparison}
\end{figure}

As primary BB observers, we selected those with sufficiently long observational periods of the best quality for each epoch.
This is illustrated in Fig. \ref{fig:schmidtwolfercomparison}, which shows the difference matrices for Wolf and Schmidt for two cases:
 Schmidt is considered as the primary observer and Wolf as the secondary (panel a) and vice versa (panel b).
It is apparent that Schmidt was a better quality observer and is more appropriate to be chosen as the primary BB observer.
By choosing Wolf as the BB observer, we would need to degrade Schmidt and other observers.

\addtocounter{table}{-1}
\begin{table}
	\centering
	\caption{(continued) Annual values of the proposed GSN series with the asymmetric $1\sigma$ intervals.}
	\begin{tabular}{|cccc|cccc|}
		\hline
		Year & G & $\sigma_+$&$\sigma_-$ &Year & G & $\sigma_+$&$\sigma_-$ \\
		\hline
		1979& 11.61&1.06&1.02&1995& 1.83&0.50&0.44\\ 1980& 10.50&1.00&1.04&1996& 1.05&0.38&0.30\\
		1981& 10.78&1.05&1.08&1997& 2.00&0.54&0.42\\ 1982& 8.85&0.99&0.88&1998& 5.53&0.80&0.78\\
		1983& 5.52&0.89&0.79&1999& 7.61&1.03&0.95\\ 1984& 3.56&0.70&0.59&2000& 9.51&1.12&0.96\\
		1985& 1.51&0.50&0.38&2001& 9.76&1.10&1.02\\ 1986& 1.19&0.43&0.31&2002& 9.52&1.10&1.04\\
		1987& 2.28&0.50&0.51&2003& 6.13&0.93&0.90\\ 1988& 6.65&0.93&0.84&2004& 4.15&0.74&0.72\\
		1989& 10.81&1.01&1.00&2005& 3.09&0.68&0.60\\ 1990& 10.86&1.16&1.11&2006& 1.95&0.50&0.45\\
		1991& 11.00&0.98&1.11&2007& 1.15&0.40&0.33\\ 1992& 7.46&1.04&0.84&2008& 0.69&0.35&0.24\\
		1993& 4.53&0.72&0.63&2009& 0.70&0.32&0.25\\ 1994& 2.96&0.64&0.56&2010& 2.03&0.54&0.43\\
		\hline
	\end{tabular}
\end{table}

To test whether our final series is robust against the choice of the primary BB observers, we repeated the
 same analysis for different BB combinations.
We used all possible combinations of high-quality long-lasting observers over four different intervals: (1) RGO (1900-1976),
 Koyama (1947-1984), Mt Wilson (1923-1958), (2) Wolfer (1880-1928), Quimby (1889-1921), (3) Schmidt (1841-1883),
 Spoerer (1861-1893), Weber (1859-1883), Wolf (1848-1893), (4) Schwabe (1826-1867), and Stark (1813-1836).
This led to 48 alternative reconstruction series.
Additionally, we constructed two more series by replacing Kanzelh\"ohe (1957-2010) with Cragg (1947-2009) and Locarno (1958-2010) and
 keeping all the other BBs as in the main series. 
Thus the total number of various GSN reconstructions was 50. 
We also included Flaugergues and Horrebow BBs in all series, but excluded the stand-alone BBs.
The reference observer was chosen between RGO, Koyama, and Mt Wilson.
Locarno has been excluded from all composites and our main series, however, we include it here as a BB to evaluate its effects on the calibration.
We note that Quimby, as an individual observer, has overlap only with RGO, Wolf and Spoerer, while Stark has no overlap with any other BB
 observer used here.
Thus, many of these auxiliary series result from disconnected BBs and are sometimes based on poor statistics. They can be used to
 assess uncertainties related to the BB selection, but as individual series, they are much less reliable than our main composite series.
In this process, we did not exclude any other observers except those automatically rejected by the code (Sec.~\ref{s:mat}).
The selection of observers within the BBs was performed automatically and may, of course, differ from those listed in
 Tables~\ref{table:rgobackbone} - \ref{table:horrebowbackbone}.

\begin{figure*}
	\centering
	\vskip 0.7cm
	\includegraphics[width=1\linewidth]{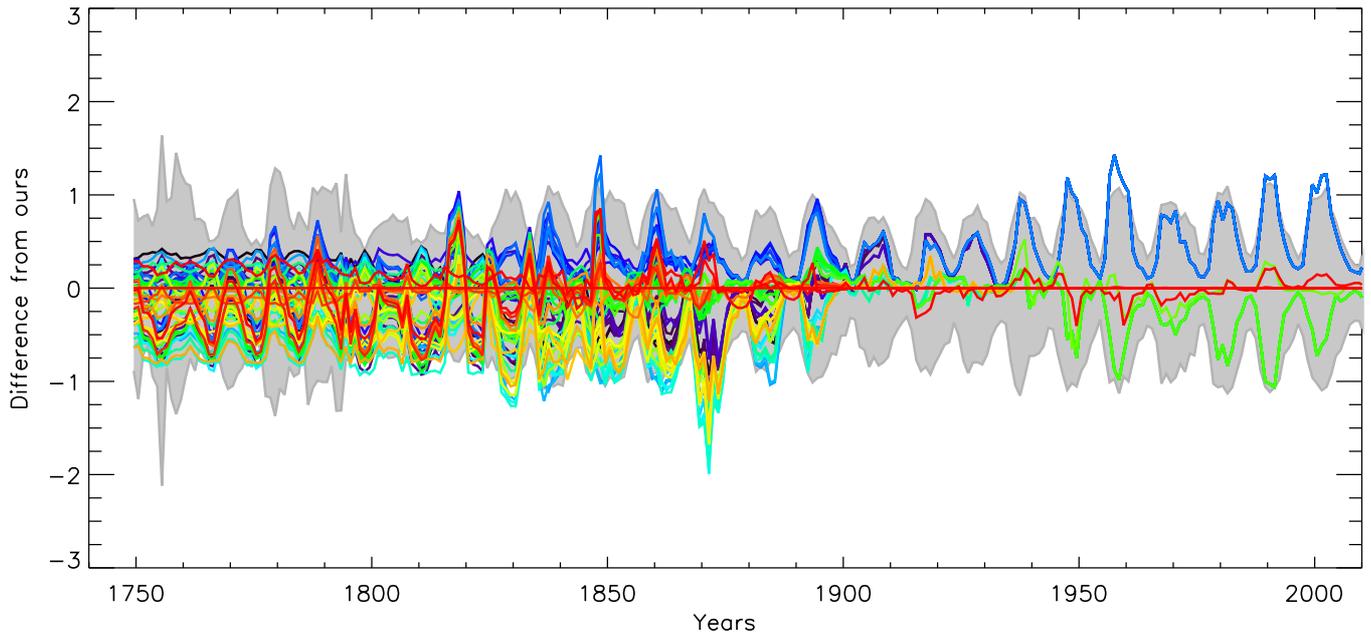}
	\caption{Difference between the main reconstructed series and all 50 auxiliary series produced with different backbone combinations. Annual values are shown.
		Grey shaded area indicates the $\pm 1\sigma$ uncertainties of the main series.}
	\label{fig:seriescomparison}
\end{figure*}

\begin{figure}
	\centering
	\includegraphics[width=1\linewidth]{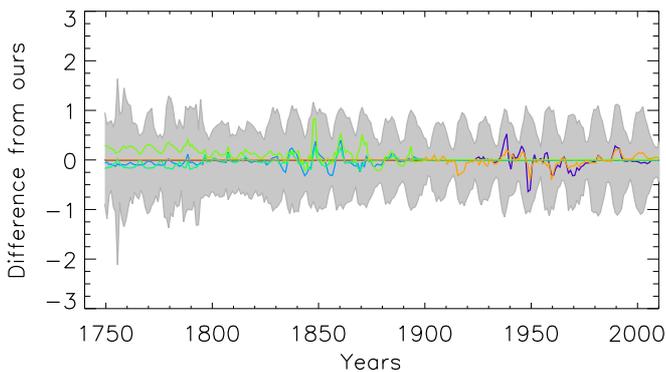}
	\caption{Difference between the main reconstructed series and the auxiliary series produced with different backbone combinations that include RGO, Wolfer, and Schwabe.
		Grey shaded area represents the $\pm 1\sigma$ uncertainties of the main series.}
	\label{fig:seriescomparison2}
\end{figure}

Figures~\ref{fig:seriescomparison} \& \ref{fig:seriescomparison2} show the differences between our main series and the different auxiliary series, described above.
The difference is mostly within the $\pm 1\sigma$ interval.
Moreover, if the three main BB observers, i.e. RGO, Wolfer, and Schwabe, are fixed, the differences among the reconstructed series are quite small (Figure~\ref{fig:seriescomparison2}) and, thus, the choice of other BBs is not important.
Using Koyama as the BB observer instead of RGO leads to systematically lower counts of sunspot groups (see blue curve in
  Figure~\ref{fig:seriescomparison}), but these counts are still within the $1\sigma$ error bars.

Thus, we can conclude that the method is stable regarding the exact choice of the BB observers with the potential
 uncertainty lying within the formal error bars.

\subsubsection{Shape of the matrix}

The majority of the calibration matrices constructed for individual observers have a shape (see Fig.~\ref{fig:WOLFERrgo})
 similar to that expected from synthetic data with an artificial acuity threshold applied \citep{usoskin_dependence_2016}.
This implies that the quality of an observer can be adequately quantified by his/her acuity observational threshold. 
However, distorted behaviour was found for some observers during periods of high solar activity, so that an observer, who is `poor'
 (counting less groups than the reference observer) during periods of low and moderate activity, may appear to report more groups
 during solar activity maxima as if he/she were a better observer than the reference observer.
This is caused by the low statistics and such columns in the matrix were replaced by the fit (Section~\ref{s:mat}).
In the case in which this behaviour occurred over an extended region of the matrix, the observers were rejected by the code.

\subsubsection{Quality of the RGO dataset}
\label{sec:rgoquality}
\begin{figure}
	\centering
	\includegraphics[width=1\linewidth]{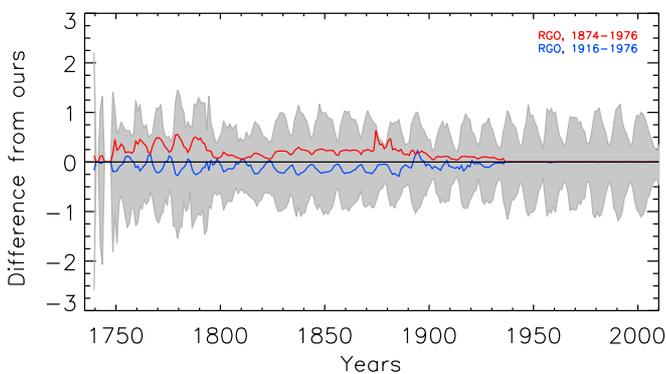}
	\caption{Differences between the main annual reconstructed $G$ series and those based on the reference RGO dataset for 1874--1976
		and for 1916--1976 (blue and red, respectively).
		The grey shaded area depicts the $\pm 1\sigma$ uncertainties of the main series.}
	\label{fig:seriescomparison_RGOdifferentperiods}
\end{figure}

We also tested how crucial the choice of the exact reference period of the RGO dataset is.
We repeated the same analysis, but considering the RGO dataset to start in 1874 and in 1916.
Since a change of the reference period affects the statistics used for the calibration,
 allocation of some individual observers to specific BBs was automatically changed and was different than in
 Tables~\ref{table:rgobackbone} through \ref{table:horrebowbackbone}.
Figure \ref{fig:seriescomparison_RGOdifferentperiods} shows the differences between the main series proposed here and
these two alternative series.
The result within the Kanzelh\"ohe BB is not affected at all,
 and for the rest of the BBs the difference is significantly smaller than the error bars, which are on average 0.14 and
  0.10 for the annual values using RGO data for the periods of 1874--1976 and 1916--1976, respectively.
At the same time, the use of the reference period shortened to after 1916 significantly decreases statistics, ignoring 42 years of RGO data.
Thus, we conclude that the present reconstruction is also robust against the choice of the reference period
 of the RGO dataset.

\subsubsection{Other issues}
Our method may suffer from an intrinsic problem related to a possible overestimate of $G$ for periods of low activity.
If a secondary `poor' observer reports no spots, the method corrects it to a finite non-zero value of $G*$ (see e.g. Fig. \ref{fig:WOLFERrgo}).
This is different from the linear $k-$factor method (e.g. SvSc16), in which zero values of a low-quality observer are always
 translated to zero values of the high-quality reference observer.

We explicitly assume, similar to all other SN reconstructions, that the observational record of any observer is error free in the sense that they report exactly the number of sunspot groups that should be visible to them on the Sun on a given day \citep[cf.][]{spearman_proof_1904,wit_uncertainties_2016}. If this assumption were violated (e.g. weather or health conditions may temporarily reduce the acuity of the observer), the method would tend to slightly underestimate the reconstructed values at high activity levels, while overestimating the values at activity minima. However, at present there is no way to assess these kinds of errors and we have to rely on this assumption. We note that this also affects all other methods, including the linear k-factor.

We also assume (as is done in all other reconstructions) that the observational quality of an observer is constant in time.
On the other hand, if it changed over time, especially outside the calibration period,
 it may introduce some additional uncertainties in the final result.
However, in this work we cannot account for that and have to make the assumption on the constancy of the quality of the observer,
 as done by all the other reconstructions as well.

\section{Summary and conclusions}
\label{s:conclusions}

We present a new reconstruction of the number of sunspot groups since 1739, along with realistic uncertainties,
 with daily, monthly, and annual time resolutions.
The reconstruction is based on the daisy-chain normalization of individual observers via so-called `backbones'
 built up on the records of the key observers of different epochs.
In contrast to most of the previous works, based on a simple linear $k-$factor scaling
 \citep[e.g. ][]{hoyt_group_1998,clette_revisiting_2014,svalgaard_reconstruction_2016}, our reconstruction
 employs a direct non-parametric calibration of observers by linking the values during days of simultaneous observations
 \citep{usoskin_dependence_2016}.
This method is based on the assumption that the quality of the data of the various observers is maintained throughout their observing period,
which may not be well validated \citep{lockwood_assessment_2016}. This will be studied elsewhere.
We also assume, as all other methods do, that daily records of each observer are error free.
A further assumption is that the main differences between the observers is due to their different observing capabilities. 
This assumption is used merely to extrapolate for the values that are missing from the overlapping period.
Thus this method works with a minimum number of assumptions and allows for a direct comparison of two observers with different observational skills.
Uncertainties of the reconstruction were assessed using a Monte Carlo method applied to the derived
 PDFs. This approach accounts naturally for the error propagation without making additional assumptions (e.g. about the normality
 and independence of errors).
In other words, we present a highly advanced daisy-chain reconstruction of GSN based on the most direct calibration of observers.

We tested the sensitivity of the method to the choice of the BB observers and of the reference period. We found
 that the reconstruction was robust and the result remained within the provided uncertainties.

The new series has been compared with other published GSN reconstructions, i.e. HoSc98, ClLi16, SvSc16, and UEA16.
The new series lies close to UEA16, but is slightly higher than that in the 18th century.
In contrast, it is systematically lower than ClLi16 in the 19th century and lower than SvSc16 in the
 18th century.
The latter two series are based on the $k-$factor scaling, which is shown to overestimate solar activity during
 solar cycle maxima \citep{lockwood_tests_2016-1,usoskin_dependence_2016,usoskin_new_2016}.
The new series confirms the existence of the modern grand maximum of activity in the second half of the 20th century,
 when sunspot cycles were significantly higher than during the 19th and 18th centuries.

The new GSN series provides a robust reconstruction of solar activity (the number of sunspot groups) with a realistic
 estimate of uncertainties and forms a basis for further investigation of centennial variability of solar activity
 over the last 270 years.

\begin{acknowledgements}
We thank the anonymous referee for useful comments.
T.C. acknowledges a postgraduate fellowship of the International Max Planck Research School on Physical Processes in the Solar System and Beyond.
This work was performed in the framework of the ReSoLVE Center of Excellence (the Academy of Finland, project no. 272157). This work was partly supported by the BK21 plus programme through the National Research Foundation (NRF) funded by the Ministry of Education of Korea.
\end{acknowledgements}

\bibliographystyle{aa}

\appendix
\section{List of observers}
In this section we list all observers that were used in each BB series. The tables contain information on the Id of the observer in the \cite{vaquero_revised_2016} database, the name of the observer,
the first year of observations employed here, the last year of observations employed here, the number of daily observations $N_{\rm d}$ used, and the number of overlap days of observations with
the BB observer $M_{\rm d}$ (for Schwabe, Flaugergues, and Horrebow BBs, the values for $\pm1$ days are also given). The BB observer is listed first and the others are sorted based on their Id.
\begin{table*}
		\centering
		\caption{List of observers used for the RGO backbone.}
	\label{table:rgobackbone}
	\begin{tabular}{llcccc}
		\hline
		Id & Observer & Start & End & $\bf{N_{\rm d}}$ & $\bf{M_{\rm d}}$\\
		\hline
		\small
		332 &  RGO										& 1900  & 1976  & 28124 & \\
		341 & Winkler, Jena & 1882  & 1910  &   6161 &   2480\\
		345 & Konkoly, Ogyalla & 1885  & 1905  &   3531 &    965\\
		347 & Stonyhurst College Obs. & 1886  & 1935  &   4534 &   4338\\
		352 & Quimby, Philadelphia & 1889  & 1921  &  10860 &   7428\\
		358 & Mount Holyoke College & 1890  & 1925  &   2799 &   2774\\
		361 & Schwab, Kremsmunster & 1892  & 1909  &   3619 &   2060\\
		362 & Catania & 1893  & 1918  &   7620 &   5417\\
		366 & Sykora, Charkow & 1894  & 1910  &   1883 &   1248\\
		368 & Lewitzky, Jurjew & 1895  & 1907  &   1279 &    647\\
		370 & Broger, Zurich & 1896  & 1935  &   9492 &   8600\\
		376 & Woinoff, Moscow & 1898  & 1919  &   2881 &   2758\\
		378 & Freyberg, St. Petersburg & 1898  & 1903  &    530 &    393\\
		380 & Kleiner, Zobten & 1899  & 1918  &   1965 &   1823\\
		381 & Kitschigin, Spitzbergen & 1900  & 1900  &    102 &    102\\
		382 & Subbotin, St. Petersburg & 1900  & 1908  &   1017 &   1017\\
		383 & Gorjatschy, Moscow & 1901  & 1908  &    603 &    603\\
		384 & Larionoff, Mohilew & 1901  & 1903  &    202 &    202\\
		385 & Struve, Charkow & 1901  & 1902  &    179 &    179\\
		386 & Guillaume, Lyon & 1902  & 1925  &   6340 &   6340\\
		387 & Schatkow, Kola & 1902  & 1910  &   1057 &   1057\\
		388 & Messerschmitt, Munchen & 1902  & 1910  &   1715 &   1715\\
		389 & Stempell, Hannover & 1903  & 1925  &   2760 &   2760\\
		390 & Amherst College Observatory & 1903  & 1906  &    672 &    672\\
		392 & Morosoff, Moscow & 1904  & 1909  &     58 &     58\\
		394 & Wasnetzoff, Moscow & 1905  & 1912  &    455 &    455\\
		395 & Belar, Laibach & 1906  & 1906  &    144 &    144\\
		396 & Hrase, Prague & 1906  & 1916  &   1748 &   1748\\
		397 & Brunner, Chur & 1906  & 1906  &    127 &    127\\
		398 & Bodocs, Ogyalla & 1906  & 1916  &   1674 &   1674\\
		399 & Ginori, Florence & 1907  & 1907  &    114 &    114\\
		402 & Sykora, Taschkent & 1907  & 1907  &    155 &    155\\
		403 & Biske, Zurich & 1908  & 1909  &    377 &    377\\
		405 & Lucchini, Florence & 1908  & 1914  &   1190 &   1190\\
		406 & Guerrieri, Capodimonte & 1908  & 1910  &    943 &    943\\
		407 & Braak, Batavia & 1909  & 1925  &   1586 &   1586\\
		408 & Stefko, Leysin & 1909  & 1913  &    260 &    260\\
		409 & Schwarz, Kremsmunster & 1910  & 1914  &    654 &    654\\
		411 & Kavan, Prague & 1911  & 1913  &    771 &    771\\
		412 & Moye, Montpellier & 1911  & 1925  &   4744 &   4744\\
		413 & Miloradowitsch, Pulkowo & 1913  & 1914  &    143 &    143\\
		414 & Buttlar, Simsdorf & 1914  & 1925  &   1898 &   1898\\
		417 & Bugoslawsky, Moscow & 1916  & 1918  &    411 &    411\\
		419 & Reed, Kennebunk, Maine & 1917  & 1917  &     33 &     33\\
		427 & Mt. Wilson, Full Disk & 1923  & 1958  &  11666 &  11666\\
		428 & Brunner, Zurich & 1926  & 1944  &   4901 &   4901\\
		429 & Buser, Arosa & 1928  & 1937  &   2722 &   2722\\
		431 & Brunner, W., Zurich & 1929  & 1944  &   3262 &   3262\\
		432 & N.A.O., Japan, $k$=0.75 & 1930  & 1930  &    244 &    244\\
		433 & N.A.O., Japan, $k$=0.65 & 1931  & 1934  &    920 &    920\\
		434 & N.A.O., Japan, $k$=0.70 & 1935  & 1948  &   1293 &   1293\\
		436 & Waldmeier, Zurich & 1936  & 1947  &   1615 &   1615\\
\hline
		\end{tabular} \end{table*} \addtocounter{table}{-1}
\begin{table*}
	\centering
		\caption{List of observers used for the RGO backbone (continued)}
	\begin{tabular}{llcccc} \hline
		Id & Observer & Start & End & $\bf{N_{\rm d}}$ & $\bf{M_{\rm d}}$ \\ \hline
		437 & N.A.O., Japan, $k$=0.55 & 1936  & 1936  &    207 &    207\\
		438 & Protitch, M., Belgrade & 1936  & 1954  &   3357 &   3357\\
		439 & N.A.O., Japan, $k$=0.60 & 1937  & 1944  &   2059 &   2059\\
		440 & Rapp, Locarno-monti & 1941  & 1944  &   1298 &   1298\\
		441 & Valencia Obs., Valencia & 1920  & 1956  &   5734 &   5734\\
		442 & Waldmeier, Arosa & 1942  & 1944  &    308 &    308\\
		443 & Djurkovic, P.M., Belgrade            &    1946&  1946 &    159  &  159  \\
		444 & Oskanjan, V., Belgrade                &   1947&  1949  &   331 &   331 \\
		445 & Koyama, H., Tokyo & 1947  & 1996  &   9848 &   5746\\
		446 & U.S. Naval Observatory & 1948  & 1956  &   3211 &   3211\\
		447 & National Astron. Obs., Japan & 1949  & 1993  &  12243 &   7689\\
		448 & Simic, M., Belgrade & 1949  & 1950  &    158 &    158\\
		449 & Dizer, M., Kandilli Obs. & 1949  & 1954  &    691 &    691\\
		451 & San Miguel Obs., Argentina & 1952  & 1965  &   1274 &   1274\\
		452 & Ozguc, A., Kandilli Obs. & 1955  & 1968  &   1931 &   1931\\
		454 & Rome Observatory & 1958  & 1989  &   7104 &   4758\\
		458 & Dogan, N., Ankara & 1974  & 1975  &    455 &    455\\
		464 & Luft, H. & 1924  & 1988  &  10628 &   7536\\
		486 & Athenes Eugenides, Greece & 1967  & 1982  &   2386 &   1877\\
		493 & Athenes III, Elias, Greece & 1949  & 1995  &   7611 &   4441\\
		610 & Luft 2, U.S.A. & 1958  & 1988  &   4992 &   2662\\
		612 & Looks, Chile & 1967  & 1987  &   3678 &   1906\\
		655 & Potsdam, Germany & 1950  & 1999  &   5436 &   2740\\
		658 & Quezon, Philippines & 1957  & 2010  &  10606 &   3709\\
		667 & Roma 3, Italy & 1950  & 2000  &   4213 &    654\\
		671 & Santiago, Chile & 1957  & 2005  &   3781 &   1356\\
		679 & Skalnate, Slovakia & 1950  & 2010  &   9200 &   4379\\
		681 & San Miguel, Argentina & 1967  & 2010  &   9400 &   2402\\
		701 & Uccle, Belgium & 1949  & 2010  &  13283 &   5033\\
		736 & Cragg, T., Los Angeles & 1947  & 2009  &  17726 &   8900\\
		\hline
\end{tabular}
\end{table*}

\begin{table*}
	\centering
	\caption{Same as Table~\ref{table:rgobackbone} but for the Kanzelh\"ohe backbone.}
\label{table:kanzelhohebackbone}
\begin{tabular}{llcccc}
	\hline
	\small
Id & Observer & Start & End & $\bf{N_{\rm d}}$ & $\bf{M_{\rm d}}$ \\
\hline
606 & Kanzelh\"ohe Treffen, Austria & 1957	& 2010	& 12862 & \\
435 & Madrid Observatory, Madrid & 1935  & 1986  &  11931 &   3453\\
453 & Lee Observatory, Bierut & 1956  & 1975  &   6532 &   3251\\
459 & Space Environment Laboratory & 1977  & 1995  &   6922 &   4764\\
460 & Debrechen Heliophysical Obs. & 1977  & 1977  &    365 &    268\\
461 & Catania Observatory & 1978  & 1987  &   3288 &   2055\\
462 & Air Force Network & 1981  & 1991  &   3572 &   2623\\
463 & British Astron. Assoc. & 1992  & 1995  &   1002 &    806\\
470 & N.O.A.A., U.S.A. & 1983  & 1994  &   2713 &   2071\\
472 & Astr. Centre Ardenne, Belgium & 1992  & 2003  &   1220 &    955\\
473 & Andries Son, Belgium & 2003  & 2010  &   1187 &    958\\
474 & Antares, Italy & 1994  & 1995  &    170 &    145\\
476 & Aguilar, Valencia, Spain & 1985  & 1988  &    967 &    729\\
477 & Ahnert, Germany & 1981  & 1988  &   1244 &    975\\
478 & Andrew Johnston, Australia & 2009  & 2010  &    221 &    168\\
479 & Alcober Valencia Spain & 1985  & 1990  &   1177 &    896\\
481 & Ankara, Turkey & 1977  & 1990  &   2898 &   2074\\
483 & Philippe Wittelsheim, France & 1989  & 2010  &   3984 &   3255\\
487 & Australian Obs. Coonabarabran, Australia & 1988  & 2007  &   5717 &   4325\\
488 & G.O.A.S., Argentina & 1987  & 1993  &    563 &    421\\
489 & Observ. Paul Ahnert, Cottbus, Germany & 1992  & 2010  &   4463 &   3431\\
490 & Donostia, Spain & 1991  & 1993  &    225 &    188\\
491 & Athenes Nat. Obser. (1) 127, Greece & 1981  & 1998  &   4247 &   3218\\
492 & Athenes Nat. Obser. (2) 109, Greece & 1981  & 1999  &   4391 &   3303\\
494 & A4 Sanvito 32404, Italy & 1986  & 2010  &   5971 &   4642\\
495 & Balseiro, Uruguay & 1983  & 1985  &    333 &    250\\
499 & Obs.Jordano Dimitrovgrad, Bulgaria & 1995  & 2005  &   1107 &    835\\
500 & Bullon, Valencia, Spain & 1982  & 2010  &   5225 &   4083\\
501 & Bortolotti Mauro, Italia & 1997  & 2009  &   3695 &   2989\\
502 & Boscat Michael, Ca & 2008  & 2010  &    466 &    397\\
504 & Basrah, Iraq & 1986  & 1986  &    228 &    168\\
505 & Broxton Tony, U.K. & 2008  & 2010  &    625 &    508\\
506 & Bucharest, Romania & 1981  & 1998  &   3828 &   2940\\
507 & Bob Vanslooten, Netherlands & 2009  & 2010  &    294 &    227\\
509 & Beyazit Obser., Turkey & 1981  & 1998  &   4532 &   3374\\
512 & Courdurie Marcq En Baroeul, France & 1989  & 2010  &   3516 &   2670\\
515 & Claeys Vedrin, Belgium & 1988  & 2010  &   5334 &   4169\\
518 & Capricorno, Campinas, Brazil & 1981  & 2009  &   3064 &   2233\\
521 & Hans Coeckelberghs, Belgium & 2006  & 2010  &    390 &    339\\
522 & Fernandez Ruis, Santander, Spain & 1992  & 2010  &   4059 &   3215\\
523 & Culgoora Narrabri, Australia & 1985  & 2010  &   4528 &   3484\\
524 & De Backer Boom, Belgium & 1983  & 2010  &   5485 &   4325\\
527 & Deman, Belgium & 1986  & 2010  &    568 &    471\\
529 & Desrues, France & 1981  & 1985  &   1289 &    933\\
530 & Dubois Langemark, Belgium & 1985  & 2010  &   6545 &   5071\\
533 & Vasquez Carlos, Argentina & 1991  & 2000  &    776 &    581\\
534 & Ebro, Roquetes, Spain & 1949  & 2010  &  16266 &  10698\\
536 & Eleizalde, Caracas, Venezuela & 1989  & 1999  &   3159 &   2411\\
\hline
\end{tabular} \end{table*} \addtocounter{table}{-1} \begin{table*}
	\centering
\caption{List of observers used for the Kanzelh\"ohe Backbone (continued)}
\begin{tabular}{llcccc} \hline
Id & Observer & Start & End & $\bf{N_{\rm d}}$ & $\bf{M_{\rm d}}$ \\ \hline
436 & Waldmeier, Zurich & 1936  & 1947  &   1615 &   1615\\
548 & Observatory Frantiska, Czech Republic & 1997  & 2010  &   1657 &   1402\\
549 & Stefaniks, Obs. Prague, Czech Republic & 1997  & 2010  &   1551 &   1308\\
550 & Fujimori Nagano, Japan & 1968  & 2010  &  10558 &   7724\\
552 & Gema Araujo, Spain & 2000  & 2010  &   3105 &   2494\\
553 & Andre Gabriel, Belgium & 2006  & 2010  &   1497 &   1249\\
554 & Grognard, Belgium & 1981  & 1991  &    572 &    396\\
555 & Gerard Dinant, Belgium & 1981  & 2007  &   5031 &   3867\\
557 & Gillissen, Belgium & 1981  & 1993  &   2543 &   1925\\
558 & German Morales, Cochabamba, Bolivia & 1995  & 2010  &   4534 &   3530\\
560 & Gollkowsky Rudolstadt, Germany & 1982  & 1997  &    874 &    711\\
562 & Schott Lutz, Gerd, Germany & 2001  & 2010  &   2259 &   1839\\
563 & Guillery Pulligny, France & 1985  & 2005  &   2914 &   2395\\
565 & Huancayo, Peru & 1983  & 2006  &   1093 &    830\\
566 & Hardie Jordanstown, N.Ireland & 1989  & 1999  &   2427 &   1825\\
567 & Hancharia, Italy & 1995  & 1998  &    434 &    356\\
568 & Helwan, Egypt & 1967  & 2010  &   9743 &   6914\\
571 & Mahmoud S, Mosque Society, Egypt & 1995  & 2005  &    942 &    691\\
572 & Holloman, U.S.a. & 1983  & 2010  &   7498 &   5697\\
573 & Hvezdaren Presov, Slovakia & 1994  & 2010  &   3749 &   3013\\
576 & Hazel Collett, United-kingdom & 2003  & 2007  &    779 &    624\\
577 & Hurbanovo, Slovakia & 1969  & 2010  &   7859 &   6386\\
578 & Hvezdaren Kysucke, Slovakia & 1993  & 2010  &   4290 &   3414\\
581 & Iskum, Budapest, Hungary & 1989  & 1999  &    655 &    553\\
582 & Iseo, Italy & 1994  & 2005  &   1628 &   1389\\
583 & Jambol, Bulgaria & 1991  & 2003  &    698 &    532\\
584 & Astro. De Reux Ciney, Belgium & 1992  & 2010  &   3363 &   2647\\
585 & Jef Claes, Belgium & 2006  & 2010  &    799 &    654\\
586 & Dragesco Jean, France & 2002  & 2005  &    774 &    599\\
587 & Jahn Jost, West-Germany & 1987  & 1993  &    628 &    485\\
588 & Observatory Haskovo, Bulgaria & 1998  & 2001  &    240 &    186\\
589 & Jorge Luis Garcia, Spain & 1996  & 2010  &   1166 &    936\\
591 & Johnston Gwynedd, England & 1991  & 2009  &   3267 &   2486\\
592 & Havana Solar Station, Cuba & 2001  & 2010  &   2582 &   2057\\
595 & Jeffrey Carels, Belgium & 2006  & 2010  &   1027 &    874\\
596 & Kawaguchi, Japan & 1981  & 2010  &   8122 &   6151\\
597 & Kandilli, Turkey & 1950  & 2010  &  11250 &   7889\\
598 & Karjali, Bulgaria & 1992  & 1999  &    552 &    436\\
599 & Kladno, Czech Republic & 1993  & 2008  &   3507 &   2855\\
600 & Koyama, Japan & 1981  & 1996  &   3250 &   2401\\
601 & Observatory Rokycany, Czech Republic & 1997  & 2001  &    351 &    291\\
602 & Kislovodsk, Russia & 1981  & 2010  &   9069 &   6880\\
607 & Larguier, France & 1985  & 1994  &   2274 &   1754\\
613 & Lieve Meeus, Belgium & 2005  & 2010  &    909 &    766\\
615 & Learmouth, Australia & 1983  & 2010  &   7466 &   5614\\
616 & Larissa Observatory, Greece & 1989  & 2010  &   4751 &   3837\\
617 & Lunping, Republic Of China & 1981  & 1998  &   2965 &   2279\\
618 & Manila, Philippines & 1971  & 1988  &   5103 &   3562\\
620 & Mac Kenzie, Dover, United-kingdom & 1981  & 2010  &   8389 &   6421\\ \hline
\end{tabular}
\end{table*} \addtocounter{table}{-1} \begin{table*}
	\centering
\caption{List of observers used for the Kanzelh\"ohe Backbone (continued)}
\begin{tabular}{llcccc} \hline
Id & Observer & Start & End & $\bf{N_{\rm d}}$ & $\bf{M_{\rm d}}$ \\ \hline
621 & Madrid, Spain & 1978  & 1986  &   1036 &    734\\
622 & Meadows Peter, U.K. & 2008  & 2010  &    566 &    478\\
625 & Michaux, Belgium & 1986  & 1990  &    319 &    251\\
626 & Murmansk, Russia & 1994  & 2010  &   3041 &   2431\\
627 & Milano, Italy & 1994  & 2010  &   1805 &   1505\\
629 & Roberto De Manzano, Italy & 2003  & 2010  &   1984 &   1684\\
630 & Mochizuki Urawa, Saitama, Japan & 1978  & 2010  &   8007 &   5984\\
631 & Mira Grimbergen, Belgium & 1987  & 2010  &   2193 &   1719\\
632 & Smolyan, Bulgaria & 1990  & 2008  &    856 &    673\\
634 & Juri Gagarin, Eilenburg, Germany & 1992  & 2010  &   1818 &   1428\\
636 & Obs. Copernicus, Varna, Bulgaria & 1995  & 2002  &    494 &    352\\
639 & Nijmegen, Netherlands & 1983  & 2010  &   5344 &   4168\\
640 & Barnes, Auckland, New-zealand & 1985  & 2010  &   4037 &   3089\\
642 & Obs. Solar Bernard Lyot, Brazil & 1995  & 1996  &    178 &    121\\
645 & O.M.A. Americana, Brasil & 1987  & 1994  &    802 &    601\\
646 & Ondrejov Observ., Czech-republic & 1991  & 2010  &   4711 &   3890\\
645 & O.M.A. Americana, Brasil & 1987  & 1994  &    802 &    601\\
646 & Ondrejov Observ., Czech-republic & 1991  & 2010  &   4711 &   3890\\
649 & Vlasim, Czech Republic & 1989  & 1992  &    436 &    360\\
650 & Palehua, Hawai & 1983  & 1997  &   3512 &   2637\\
651 & Perroni, Brazil & 1981  & 1986  &   1413 &   1021\\
652 & Pasternak, Berlin, Germany & 1984  & 2010  &   5429 &   4331\\
654 & Lormont, France & 1991  & 1997  &    691 &    555\\
656 & Observatory Prostejov, Czech Republic & 1998  & 2010  &   1529 &   1273\\
657 & Pyong Yang, Korea & 1985  & 2003  &   4324 &   3306\\
659 & Ramey, Puerto-rico & 1983  & 2003  &   5957 &   4505\\
666 & Rokycany - Luzicka, Czech Republic & 1997  & 2001  &    424 &    348\\
668 & Paulo Roberto Moser, Brazil & 2010  & 2010  &    172 &    144\\
669 & Rasson Mons, Belgium & 1988  & 1997  &   2126 &   1626\\
670 & Rodriguez, Venezuela & 1986  & 1989  &    950 &    720\\
672 & Siracusa II, Lapichino, Italia & 1986  & 1995  &    365 &    294\\
673 & Sjoerd Dufoer, Belgium & 2007  & 2010  &    366 &    323\\
674 & Sergio Fabiani, Bolivia & 1995  & 1995  &    133 &    105\\
675 & Sigma Octante, Cochabamba, Bolivia & 1981  & 2010  &   5258 &   4049\\
677 & Smith Marlyn, U.K. & 2008  & 2010  &    379 &    313\\
678 & San Jose, Buenos Aires, Argentina & 1986  & 1996  &    702 &    531\\
683 & Sobota, Slovakia & 1992  & 2010  &   5258 &   4280\\
685 & Saudi Arabia, Jeddah & 1981  & 2010  &   5477 &   4154\\
688 & Suzuki, Japan & 1981  & 2010  &   7839 &   5954\\
691 & Trento, Italy & 1994  & 1994  &     48 &     48\\
692 & Thomas Teague, United Kingdom & 2005  & 2010  &    219 &    168\\
693 & Central Weather Bureau, Republic Of China & 1981  & 2010  &   5898 &   4564\\
694 & Tangjungsari, Indonesia & 1984  & 1989  &   1358 &   1031\\
696 & Taipei 2, Republic Of China & 1981  & 2005  &   3692 &   2796\\
697 & Trieste, Italy & 1967  & 1993  &   2704 &   2074\\
698 & Spaninks Tilburg, Netherlands & 1991  & 2010  &   3079 &   2424\\
700 & Tony Tanti Naxxar, Malta & 1986  & 1998  &   2271 &   1769\\
702 & U.L.B., Belgium & 1983  & 1986  &    594 &    463\\
705 & Sliven, Bulgaria & 1989  & 2003  &   1301 &    985\\
706 & Ventura Mosta, Malta & 1986  & 2003  &   3732 &   2834\\
709 & Ruben Verboven, Belgium & 2006  & 2010  &    154 &    135\\
713 & Monte Mor, Brazil & 2006  & 2010  &    729 &    625\\
717 & Y Alarcos, Valencia, Spain & 1986  & 1994  &    587 &    441\\
719 & Yvergneaux Ronse-renaix, Belgium & 1981  & 1997  &   3754 &   2844\\
720 & Zagora, Bulgaria & 1990  & 2010  &   2851 &   2307\\
721 & Zamora, Spain & 1993  & 1999  &   1138 &    865\\
		\hline
	\end{tabular}
\end{table*}

\begin{table*}
		\centering
		\caption{Same as Table~\ref{table:rgobackbone} but for the Wolfer backbone.}
	\label{table:wolferbackbone}
	\begin{tabular}{llcccc}
		\hline
			\small
			Id & Observer & Start & End & $\bf{N_{\rm d}}$ & $\bf{M_{\rm d}}$ \\
		\hline
		335+338 & Wolfer, Zurich	   & 1876  & 1928  &  13533 &  \\
		329 & Secchi, Rome & 1871  & 1877  &   1530 &    298\\
		333 & Moncalieri & 1874  & 1893  &   3598 &   2422\\
		336 & Aguilar, Madrid & 1876  & 1882  &   1940 &   1381\\
		337 & Monthly Weather Review & 1877  & 1886  &   2383 &   1786\\
		339 & Ricco, Palermo & 1880  & 1892  &   3709 &   2668\\
		343 & Merino, Madrid & 1883  & 1896  &   3221 &   2394\\
		346 & Vogel, Potsdam & 1886  & 1886  &    162 &    135\\
		347 & Stonyhurst College Obs. & 1886  & 1935  &   4534 &   1835\\
		349 & Schmoll, Paris & 1888  & 1892  &   1359 &   1041\\
		350 & Haverford College Obs., PA & 1888  & 1899  &   2063 &   1547\\
		353 & Carleton College Observatory & 1889  & 1892  &    523 &    383\\
		355 & Smith Observatory & 1890  & 1891  &    258 &    192\\
		356 & Hadden, D.E., Alta, Iowa & 1890  & 1890  &   2964 &   2256\\
		359 & Schreiber, Kalocsa & 1891  & 1895  &   1173 &    976\\
		360 & Zona, Palermo & 1891  & 1891  &    282 &    233\\
		369 & Maier, Schaufling & 1895  & 1901  &    632 &    529\\
		373 & Oliver, A.I., Boston U., MA & 1897  & 1901  &    254 &    190\\
		375 & Jastremsky, B., Charkow & 1898  & 1900  &    149 &    111\\
		377 & Mirkowitsch, Jaroslaw & 1898  & 1900  &    135 &    111\\
		379 & Kaulbars, St. Petersburg & 1898  & 1901  &    649 &    508\\
		391 & Boston University Obs. & 1903  & 1906  &    359 &    239\\
		401 & Bemmelen, Batavia & 1907  & 1919  &   2748 &   1910\\
		415 & Schmid, St. Gallen & 1915  & 1915  &    225 &    173\\
		421 & Voss, Altona & 1918  & 1918  &    198 &    145\\
		465 & Wolf, R., Zurich (small Telescope) & 1858  & 1893  &   8285 &   4385\\
		\hline
			\end{tabular}
		\end{table*}

\begin{table*}
		\centering
		\caption{Same as Table~\ref{table:rgobackbone} but for the  Schmidt backbone.}
	\label{table:schmidtbackbone}
	\begin{tabular}{llcccc}
		\small
		Id & Observer & Start & End & $\bf{N_{\rm d}}$ & $\bf{M_{\rm d}}$ \\
		\hline
	292 & Schmidt, Athens  & 1841  & 1883  & 6970  &  \\			
	298 & Wolf, R., Zurich & 1848  & 1893  &  18311 &   4153\\
	307 & Carrington, London & 1853  & 1860  &   1215 &    204\\
	311 & Weber, Peckeloh & 1859  & 1883  &   6983 &   4035\\
	318 & Spoerer, G., Anclam & 1861  & 1893  &   6281 &   2449\\
	323 & Ferrari, Rome & 1866  & 1879  &    478 &    429\\
	324 & Leppig, Leipzig & 1867  & 1881  &   2611 &   1979\\
	325 & Dawson, W.M., Spiceland, Ind & 1867  & 1890  &   1623 &    824\\
	328 & Tacchini, Rome & 1871  & 1900  &   7584 &   2388\\
	330 & Billwiller, Zurich & 1872  & 1875  &    308 &    286\\
	331 & Sawyer, E.F., Cambridgeport & 1872  & 1874  &    282 &    273\\
	342 & Janesch, Laibach & 1882  & 1887  &   1164 &    439\\
					\hline
\end{tabular}
\end{table*}

\begin{table*}
		\centering
		\caption{Same as Table~\ref{table:rgobackbone} but for the  Schwabe backbone.}
	\label{table:schwabebackbone}
	\begin{tabular}{llccccc}
				\hline
				\small
		Id & Observer & Start & End & $\bf{N_{\rm d}}$ & $\bf{M_{\rm d}}$ & $\bf{M_{\rm d}} \pm$1day \\
		\hline
		279 & Schwabe, H. Dessau    & 1825  & 1867  &  11945 &  11945 & \\
		255 & Stark, J.M., Augsburg & 1826  & 1836  &   1075 &   924 & 1029\\
		274 & Herschel, J., London & 1822  & 1837  &    122 &     37 & 61\\
		278 & Von Both, G., Breslau & 1825  & 1826  &    183 &     59 &72\\
		280 & Hussey, T.J., England & 1826  & 1837  &   1207 &   879 & 1073\\
		282 & Lawson, H., Hereford & 1831  & 1832  &    200 &    151 & 180\\
		283 & Ruprecht, H., Ziegenhain & 1832  & 1832  &     39 &     31 & 35\\
		284 & Boguslawski, P.H.L., Breslau & 1832  & 1832  &     17 &     14 & 17\\
		285 & Bohm, J.G., Wien & 1833  & 1836  &    101 &     84 & 96\\
		290 & Petersen, A.C., Altona & 1840  & 1841  &     13 &     10 & 13\\
		294 & Peters, C.H.F., Clinton, NY & 1844  & 1870  &   1308 &  953 &  1028\\
		299 & Greisbach, T.J., England & 1850  & 1865  &    168 &   161 & 168\\
		300 & Sestini, Georgetown & 1850  & 1850  &     42 &     35 & 39\\
		304 & Pogson, N., London & 1851  & 1851  &     13 &     11 & 13\\
		305 & Tomaschek, Wien & 1852  & 1854  &     15 &     8 & 15\\
		306 & Borck, Cassel & 1852  & 1855  &     19 &     19 & 19\\
		308 & Flagstaff Obs., Melbourne & 1857  & 1858  &     16 &     15 & 16\\
		312 & Howlett, F., England & 1859  & 1892  &    766 &    505 & 537\\
		313 & Baxendall, J., Manchester & 1859  & 1859  &      7 &     7 & 7\\
		314 & Coast Survey, Washington & 1860  & 1862  &    475 &   430 & 460\\
		316 & Jenzer, Bern & 1861  & 1865  &    585 &    542 & 566\\
		320 & Waldner, Zurich & 1863  & 1864  &     41 &     39 & 41\\
		321 & Meyer, Zurich & 1864  & 1871  &    912 &    387 & 397\\
						\hline
\end{tabular}
\end{table*}

\begin{table*}
		\centering
	\caption{Same as Table~\ref{table:rgobackbone} but for the Flaugergues backbone.}
		\label{table:flaugerguesbackbone}
		\begin{tabular}{llccccc} \hline
				\small
		Id & Observer & Start & End & $\bf{N_{\rm d}}$ & $\bf{M_{\rm d}}$ & $\bf{M_{\rm d}} \pm$1day \\ \hline
		22+227 & Flaugergues, H., Aubenas and Viviers                & 1788 & 1830 &   2101   &2101& \\
		202 & Bode, J.E., Berlin & 1774  & 1822  &     68 &     26& 32\\
		218 & Heinrich, P., Munich & 1781  & 1820  &    396 &    119 &216\\
		236 & Herschel, W., London & 1794  & 1818  &    384 &     29 & 67\\
		238 & Gemeiner, A.T., Regensburg & 1797  & 1797  &      3 &  1&    3\\
		245 & Lindener, B.A., Glatz & 1800  & 1827  &    519 &   114 & 210\\
		246 & Derfflinger, T., Kremsmunster & 1802  & 1824  &    789 &   47 & 101\\
		250 & Prantner, S.M.J., Wilten & 1804  & 1844  &    115 &   35 &  67\\
		258 & Tevel, C., Middelburg & 1816 & 1836 & 858 & 89 & 156 \\
		260 & Watts, Cape Diamond, Quebec & 1816  & 1818  &     83 &    3&  10\\
		262 & Adams, C.H., Edmonton & 1819  & 1823  &    977 &   34&  66\\
		263 & Pastorff, J.W., Drossen & 1819  & 1833  &   1477 &   53& 109\\
		273 & Arago, F.D., Paris & 1822  & 1830  &    923 &   85& 145\\
						\hline
\end{tabular}
\end{table*}

\begin{table*}
		\centering
	\caption{Same as Table~\ref{table:rgobackbone} but for the Horrebow backbone.}
\label{table:horrebowbackbone}
	\begin{tabular}{llccccc} \hline
				\small
		Id & Observer & Start & End & $\bf{N_{\rm d}}$ & $\bf{M_{\rm d}}$ & $\bf{M_{\rm d}} \pm$1day \\ \hline
180 & Horrebow, C., Copenhagen           &      1761 & 1776 &   1532 & 1532&\\
174 & Lalande, J., Paris & 1752  & 1798  &    105 &     15 &26\\
185 & Warschauer & 1764  & 1766  &      3 &     2& 3\\
203 & Lievog, E., Copenhagen & 1776  & 1777  &    196 &   97& 101\\
466 & Staudach, J.C., Nuremberg & 1749  & 1799  &   1172 &   128 & 234\\
\hline
\end{tabular}
\end{table*}
\end{document}